\documentclass[letterpaper,twocolumn,10pt]{article}
\usepackage{usenix2019_v3}
\usepackage{comment}

\usepackage{tabularx}
\usepackage{graphicx}
\usepackage{xspace}
\usepackage{xcolor}
\usepackage{listings}
\usepackage{color}
\usepackage{subfig}
\usepackage{fancyhdr}

\DeclareFixedFont{\ttb}{T1}{txtt}{bx}{n}{9} 
\DeclareFixedFont{\ttm}{T1}{txtt}{m}{n}{9}  

\definecolor{deepblue}{rgb}{0,0,0.5}
\definecolor{deepred}{rgb}{0.6,0,0}
\definecolor{deepgreen}{rgb}{0,0.5,0}

\newcommand\pythonstyle{\lstset{
language=Python,
basicstyle=\ttm,
numbers=left,
otherkeywords={self},             
keywordstyle=\ttb\color{deepblue},
emph={MyClass,__init__},          
emphstyle=\ttb\color{deepred},    
stringstyle=\color{deepgreen},
frame=tb,                         
showstringspaces=false            %
}}

\definecolor{delim}{RGB}{20,105,176}
\definecolor{numb}{RGB}{106, 109, 32}
\definecolor{string}{rgb}{0.64,0.08,0.08}

\lstdefinelanguage{json}{
    numbers=left,
    numberstyle=\scriptsize,
    frame=none,
    rulecolor=\color{black},
    showspaces=false,
    showtabs=false,
    breaklines=true,
    postbreak=\raisebox{0ex}[0ex][0ex]{\ensuremath{\color{gray}\hookrightarrow\space}},
    breakatwhitespace=true,
    basicstyle=\ttfamily\scriptsize,
    upquote=true,
    morestring=[b]",
    stringstyle=\color{string},
    literate=
     *{0}{{{\color{numb}0}}}{1}
      {1}{{{\color{numb}1}}}{1}
      {2}{{{\color{numb}2}}}{1}
      {3}{{{\color{numb}3}}}{1}
      {4}{{{\color{numb}4}}}{1}
      {5}{{{\color{numb}5}}}{1}
      {6}{{{\color{numb}6}}}{1}
      {7}{{{\color{numb}7}}}{1}
      {8}{{{\color{numb}8}}}{1}
      {9}{{{\color{numb}9}}}{1}
      {\{}{{{\color{delim}{\{}}}}{1}
      {\}}{{{\color{delim}{\}}}}}{1}
      {[}{{{\color{delim}{[}}}}{1}
      {]}{{{\color{delim}{]}}}}{1},
}

\lstnewenvironment{python}[1][]
{
\pythonstyle
\lstset{#1}
}
{}

\newcommand\py[1]{{\pythonstyle\lstinline!#1!}}

\newcommand{\hstore}{\code{hstore}\xspace}
\newcommand{\hstorecc}{\code{hstore-cc}\xspace}
\newcommand{\mapstore}{\code{mapstore}\xspace}
\newcommand{\pmem}{PM\xspace}

\newcommand{\code}[1]{\begin{ttcodefont}#1\end{ttcodefont}}
\newcommand{\smallcode}[1]{\begin{ttsmallcodefont}#1\end{ttsmallcodefont}}

\definecolor{Gray}{gray}{0.85}
\definecolor{LightCyan}{rgb}{0.88,1,1}
\newcolumntype{a}{>{\columncolor{Orange}}c}
\newcolumntype{b}{>{\columncolor{white}}c}

\date{}

\title{\Large \bf An Architecture for Memory Centric Active Storage (MCAS)}

\author{
  {\rm Daniel G. Waddington}\\
  daniel.waddington@ibm.com \\
  IBM Research Almaden
\and
  {\rm Clem Dickey}\\
  dickeycl@us.ibm.com \\
  IBM Research Almaden
\and
  {\rm Moshik Hershcovitch}\\
  moshikh@il.ibm.com \\
  IBM Research Almaden
\and
  {\rm Sangeetha Seshadri}\\
  seshadrs@us.ibm.com \\
  IBM Research Almaden
} 

\begin{document}

\date{} 

\maketitle 


\begin{abstract}
  
The advent of CPU-attached persistent memory technology, such as
Intel's Optane Persistent Memory Modules (PMM), has brought with it new
opportunities for storage.  In 2018, IBM Research Almaden
began investigating and developing a new enterprise-grade storage
solution directly aimed at this emerging technology.

MCAS (Memory Centric Active Storage) defines an ``evolved''
network-attached key-value store that offers both near-data compute
and the ability to layer enterprise-grade data management services on
shared persistent memory.  As a \textit{converged memory-storage
  tier}, MCAS moves towards eliminating the
traditional separation of compute and storage, and thereby unifying
the data space.

This paper provides an in-depth review of the MCAS architecture and
implementation, as well as general performance results.

\end{abstract}


\section{Introduction}

Traditionally, the separation between volatile data in memory and
non-volatile data in storage devices (e.g., SSD) has been clear.  The
interface and semantics between the two domains is well defined; that
is, in the event of power-reset or power-fail events, data in memory
is lost and, in turn, is then retrieved from storage during the
recovery process.

\begin{figure}[ht]
\centering
\includegraphics[width=0.8\columnwidth]{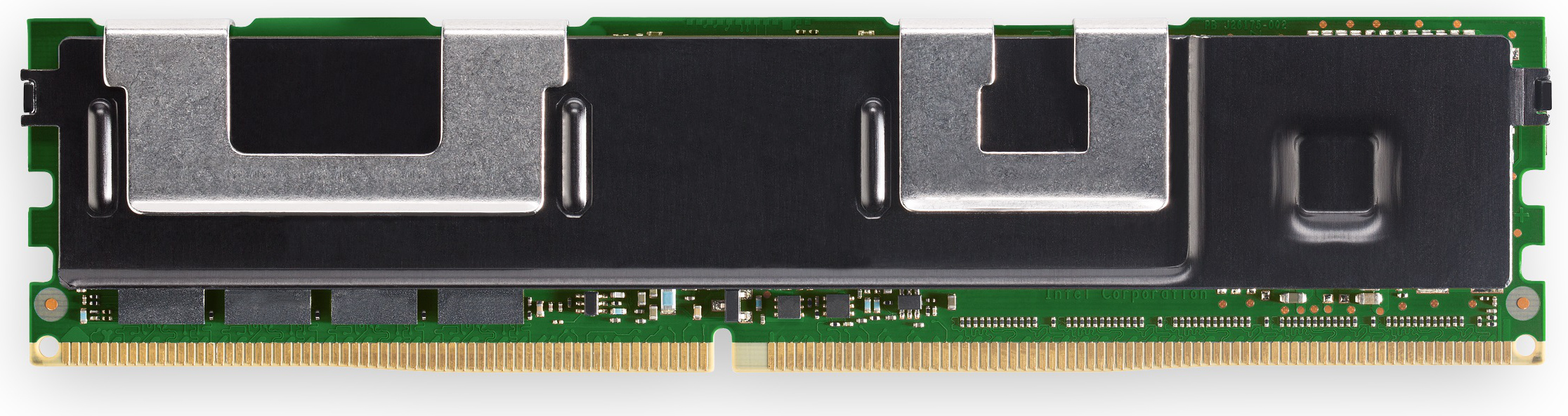}
\caption{Optane DC Persistent Memory Module}
\label{fig:aep}
\end{figure}

With the advent of Persistent Memory (herein abbreviated to \pmem),
such as Intel's Optane DC Persistent Memory Modules (see
Figure~\ref{fig:aep}), this conventional separation of memory and
storage begins to blur.  Because \pmem behaves like memory, operations
on data held within \pmem can be performed in-place without having to
first load, and potentially de-serialize, from storage.  Likewise,
data written to \pmem need not be pushed down to storage to assure its
retention.  The result is that operations on durable data can be
performed at an order-of-magnitude lower latency than has been
previously possible.  Nevertheless, the Achilles' heel of \pmem is
that data management services traditionally realized by enterprise
storage systems (e.g., access control, encryption, replication,
compression, versioning, geo-distribution) cannot be easily realized
with additional software.

\pmem raises the data preservation boundary up
the stack into the main memory space.  It provides non-volatile memory
hardware that sits below existing volatile caches but that, unlike
existing DRAM-based memory, retains data in the event of failure
or reset.  The caveat is that data must be explicitly flushed from the
CPU cache (e.g. via \code{clflushopt}) for its persistence to be guaranteed.

It is also ``byte-addressable'' in that it is directly accessed via
load-store instructions provided by the CPU.  Intel Optane DC PMM,
which uses 3D XPoint (3DXP) technology, operates at a cache-line read-write
latency of around $300ns$ (see ~\cite{spectra2020,
  izraelevitz2019basic} for more detail).  Even though this is slower
than DRAM access latencies ($\sim100ns$) it is at least 30x faster
than state-of-the-art storage (e.g., NVMe SSD).  Capacity of \pmem is
also about 8x that of DRAM\footnote{For 3DXP, which is based on
  lattice-arranged Phase Change Memory (PCM).}.

Another consequence of \pmem being attached to the system as memory
is that it allows use of Direct Memory Access (DMA) and
Remote DMA (RDMA) to move data around.  For example, data can be
copied from \pmem to the network (via RDMA) or to another device such as
a GPU (via DMA), without requiring execution by the CPU.  This frees
the CPU to perform other tasks rather than executing \code{memcpy}
loops in order to move data.  Today, NVIDIA/Mellanox RDMA network
adapters can transfer data at near 400Gbps (50GiB/s) and therefore,
using multiple adapters, can even keep pace with the performance of \pmem.

\subsection{Current Limitations of Intel Optane \pmem}

While Intel Optane PMM provides many useful \pmem features as just
discussed a number of limitations are evident in the current generation.

\begin{itemize}

\item \textit{Endurance} - lifetime endurance of the hardware is
  significantly less than DRAM (3DXP at $10^6$ writes, DRAM at $10^{10}$)
  although orders-of-magnitude higher than NAND-flash. For intensive
  data write operations pushing through the cache this may be a
  significant limitation~\cite{227810}.

\item \textit{Asymmetric Performance Scaling} - write performance does
  not scale with increasing number of threads, while read performance
  scales at around 1.2\% degradation (from linear) per-core up to 28
  cores~\cite{9238605}.

\item \textit{64-bit Aligned Atomicity} - only aligned 64-bit writes
  can be guaranteed to happen atomically by the hardware.  There is
  currently no hardware support for multi-write atomicity/transactions and
  therefore this burden is left to the software.

\item \textit{Reliability \& Serviceability} - to provide maximum
  performance DIMMs must be configured to stripe data across 6
  devices.  In the event of a single DIMM failure, data on all of the
  DIMMs is effectively lost.
  
\item \textit{Cost} - although current 4Q2020 cost is $\sim0.5$x than
  that of DRAM, it is an order-of-magnitude higher than NAND-flash
  ($\sim 7\$/GB$ versus $\sim 1\$/GB$)

\end{itemize}

\section{Design Objectives and Solution Positioning}

With the previously discussed characteristics of \pmem in mind, the following
tenets in the design of MCAS were made.  The solution should:

\begin{enumerate}

\item Allow \pmem h/w resources to be shared as a network-attached
  capability using RDMA to provide maximum transfer speed.  Data
  sharing across independent nodes should be possible with appropriate
  locking/serialization provided by the MCAS system.
\item Maintain an immediate consistency model with guaranteed
  persistence (i.e. data is known to be flushed from volatile caches when a write is made).
\item Support zero-copy (RDMA-only) transfer of large data chunks
  enabling bulk data movement with CPU \code{memcpy} execution.
\item Minimize round-trip latency so that small reads/writes can be
  performed synchronously reducing s/w complexity in the client.
\item Scale through sharding to bound any performance degradation due
  to locking.
\item Provide flexible software-defined deployment for both
  on-premises and cloud.  Support both containerized and
  virtual-machine based deployment scenarios.
\item Enable safe, in-place user-defined operations directly on \pmem
  (both general and domain-specific).
\item Provide flexibility in the custom service layering (e.g.,
  combining replication and tiering).
  
\end{enumerate}

MCAS is positioned as a \textit{converged memory-storage tier}
providing high-performance random access to durable data.  Because
MCAS is based on \pmem it can provide fine-grained durability (per
write) as opposed to snap-shotting.  Even with synchronous, consistent
and guaranteed-persistent replication across multiple nodes, MCAS can
support still millions of updates per second.


\section{MCAS Core Architecture}

\begin{figure}
\centering
\includegraphics[width=0.9\linewidth]{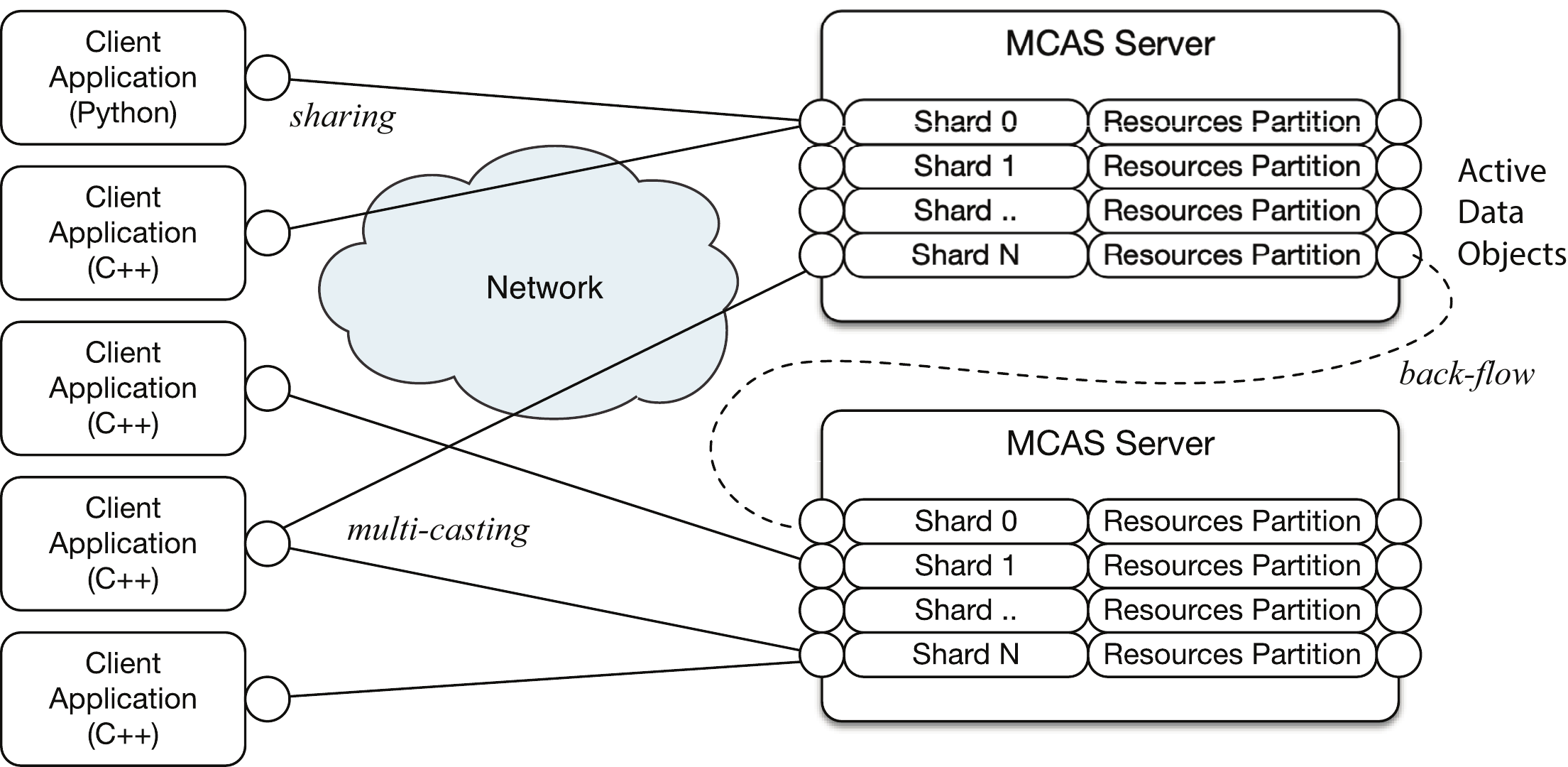}
\caption{MCAS High-Level Architecture}
\label{fig:hl_arch}
\end{figure}

MCAS is implemented as a Linux process (known as the 'shard' process)
of which multiple instances can be concurrently deployed on the same
machine.  An MCAS process instance manages one or more network
end-points, each corresponds to a separate \textit{shard} (see
Figure~\ref{fig:hl_arch}).  

Shards are single-threaded and manage request handling for a
\textit{set} of pools.  They can be accessed concurrently by multiple
clients (from different nodes in the network) and can be
grouped into larger virtual data domains through client-side
clustering techniques such as consistent
hashing~\cite{TanenbaumSteen07}.

Each shard optionally maintains Active Data Object (ADO) processes
that provide custom functionality to the store.  The ADO processes
themselves may act as clients to other MCAS nodes (back-flow).  ADOs
are discussed in more detail in Section~\ref{sec:ado}.

Resources (memory, CPU cores) are statically allocated to
each shard through the MCAS configuration file. An example two-shard
configuration file is shown in Listing~\ref{lst:conf0}.

\begin{minipage}{\linewidth}
\begin{lstlisting}[language=json,
    caption={Example MCAS two-shard configuration},
    captionpos=b, label={lst:conf0}]
{
  "shards" :
  [
    {
      "core" : 0,
      "port" : 11911,
      "net"  : "mlx5_0",
      "default_backend" : "hstore",
      "dax_config" : [{
          "path": "/dev/dax0.0",
          "addr": "0x9000000000" }]
    },
    {
      "core" : 1,
      "port" : 11912,
      "net"  : "mlx5_0",
      "default_backend" : "hstore",
      "dax_config" : [{
          "path": "/dev/dax0.1",
          "addr": "0xA000000000" }]
    }
  ],
  "net_providers" : "verbs"
}
\end{lstlisting}
\end{minipage}

Each shard serves a single network end-point established using the
\textit{libfabric} library, which is part of the Open Fabric
Interfaces (OFI)
framework\footnote{https://ofiwg.github.io/libfabric/}.  This library
provides a common abstraction layer and services for high-performance
fabrics such as RDMA verbs, Intel TrueScale, and Cisco VIC.  It also
includes a provider for plain TCP/IP socket (TCP or UDP) but without
user-level and zero-copy capabilities.  MCAS primarily supports the
RDMA verbs and sockets providers.

Pools are they next level of data collection.  Each pool can only
belong to a single shard, which in turn means that the handling of
operations for a specific key-value pair is always performed by the
same shard and thus same thread.  Pools represent the \textit{security
  boundary} from a client perspective.  That is, access control and
memory resources are all bound to a pool.  If a client has
access-rights to a pool, then they also have access-rights to all
other key-value pairs in the pool.  

\begin{figure}
\centering
\includegraphics[width=0.9\linewidth]{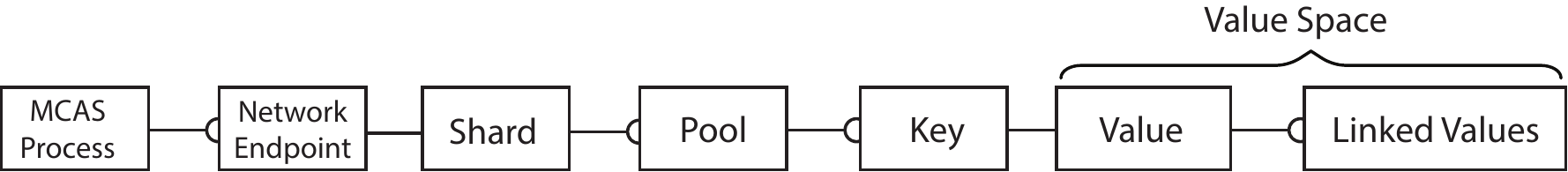}
\caption{MCAS Entity Relationships }
\label{fig:schema}
\end{figure}

The overall data entity schema is given in Figure~\ref{fig:schema}.

\subsection{Client API} \label{subsec:Client API} 

Client applications interact with MCAS by linking to the client-API
library (\code{libcomponent-mcasclient.so}).  This library provides a
C++ based interface to MCAS.  The basic operations are very typical of
a traditional key-value store (see Table~\ref{tab:clientapi}); they
operate on \textit{opaque} values that are identified by a unique
key. Both keys and values are variable length and there is no
restriction on their size.

\begin{table}[t]
\begin{centering}
\begin{tabularx}{1.0\linewidth}{
>{\setlength{\hsize}{.3\hsize}\raggedright\footnotesize}X
>{\setlength{\hsize}{.7\hsize}\raggedright\arraybackslash\footnotesize}X }
  \hline
  \textbf{Function} & \textbf{Description} \\
  \hline
  \smallcode{create\_pool} & Create a new pool or open existing pool \\
  \smallcode{open\_pool} & Open existing pool (optional create on demand) \\
  \smallcode{close\_pool} & Release handle to pool \\
  \smallcode{delete\_pool} & Securely delete pool and release pool memory to shard \\
  \hline
  \smallcode{configure\_pool} & Configure pool (e.g., add secondary index) \\
  \hline
  \smallcode{put} & Write small ($<$ 2MiB) key-value pair.  Optionally allow overwrites \\
  \smallcode{get} & Read small ($<$ 2MiB) key-value pair \\
  \smallcode{async\_put} & Asynchronous version of put \\
  \smallcode{async\_get} & Asynchronous version of get \\

  \hline
  \smallcode{free\_memory} & Free memory allocated by get call \\
  \hline
  \smallcode{erase} & Erase key-value pair from pool \\
  \smallcode{async\_erase} & Asynchronous version of erase \\
  \hline
  \smallcode{get\_attributes} & Get attributes for pool or key/value pair \\
  \smallcode{get\_statistics} & Get shard statistics \\
  \smallcode{find} & Search key space in secondary index \\  
  \hline
\end{tabularx}
\caption{Basic Client API Summary}
\label{tab:clientapi}
\end{centering}
\end{table}

\subsubsection{Zero-copy Bulk Transfers}

\begin{figure}
\centering
\includegraphics[width=1.0\linewidth]{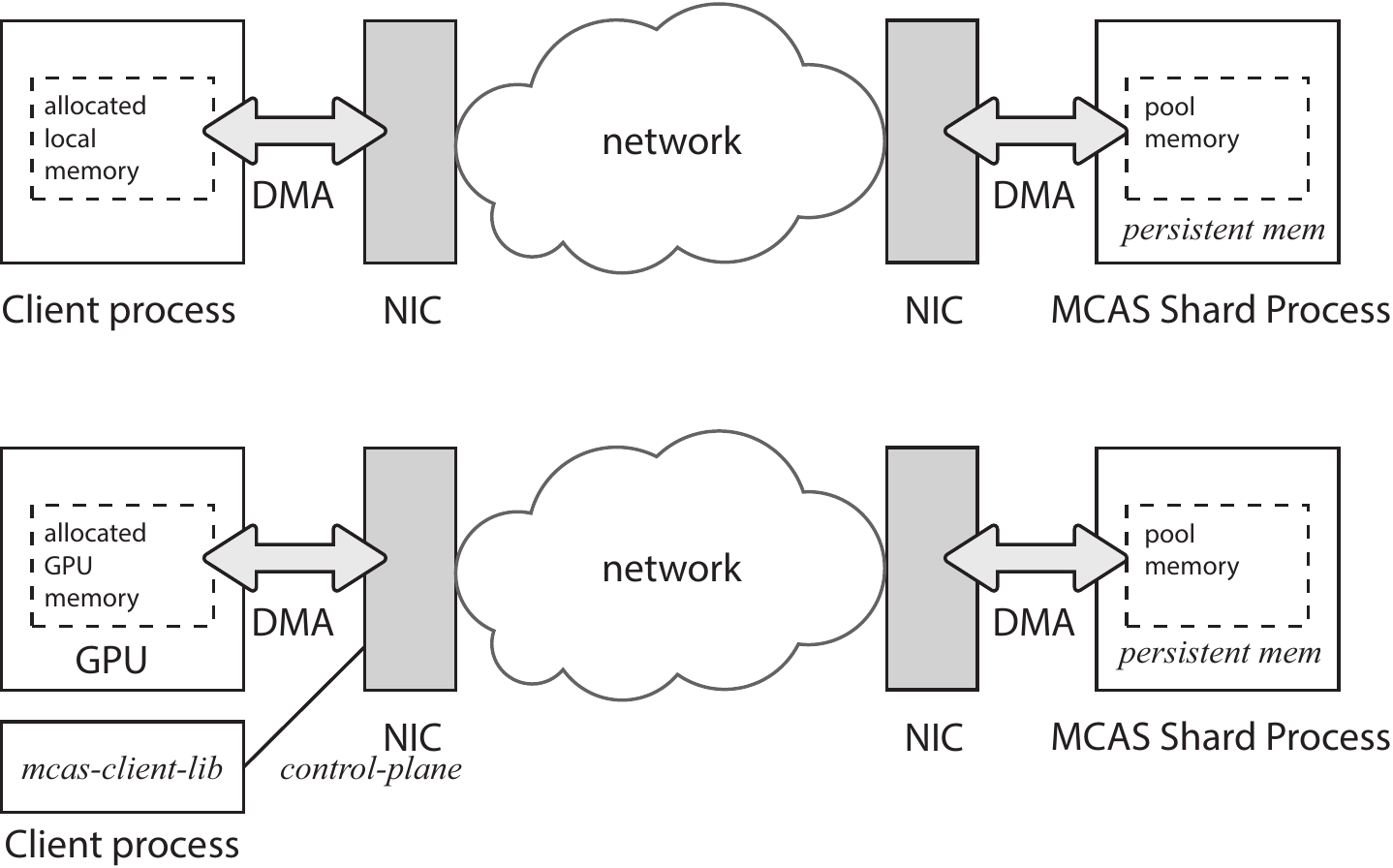}
\caption{MCAS Direct Transfers}
\label{fig:direct}
\end{figure}

MCAS also provides APIs for moving data to and from client-host memory
without a memory copy (memcpy) operation being performed under the
hood.  These \textit{direct transfer} APIs 
(see Table~\ref{tab:clientbulkapi}) are realized through the
underlying RDMA network hardware and allow data to be moved directly
from packet buffers into user-space memory (see
Figure~\ref{fig:direct}).  The memory for the direct APIs must be
allocated (e.g., via POSIX \code{alloc\_aligned}) and then registered
with the RDMA stack via the MCAS \code{register\_direct\_memory} call.
Under the hood, the direct APIs use RDMA read/write operations.
However, because the semantics of the MCAS protocol is
persistent-on-completion, two-sided operations (i.e. send/recv) are
still used to provide the acknowledgments (see Appendix A for detail).

The direct APIs can also be used with NVidia GPU-direct capabilities\footnote{https://docs.nvidia.com/cuda/gpudirect-rdma/index.html}.
This allows data to move from the MCAS server, across the network and
then directly from the NIC hardware into an application-defined region
of memory allocated inside the GPU.
In this scenario, 
the CPU
``host-code'' on the client must acquire the region of GPU memory
(e.g., via \code{cuMemAlloc}) and then register this memory with the
MCAS \code{register\_direct\_memory} call.  On completion of the
direct call on the CPU, movement of data into or out of the GPU is
known to be complete.

Depending on the PCIe arrangement and NIC hardware, direct transfers
are able to achieve transfer rates of tens of GiB/s.  Of course, these
transfers do not require CPU instruction execution and therefore the CPU is
free to perform other useful work.

\begin{table}[t]
\begin{centering}
\begin{tabularx}{1.0\linewidth}{
>{\setlength{\hsize}{.3\hsize}\raggedright\footnotesize}X
>{\setlength{\hsize}{.7\hsize}\raggedright\arraybackslash\footnotesize}X }
  \hline
  \textbf{Function} & \textbf{Description} \\
  \hline
  \smallcode{register\_direct\\\_memory} & Register client-allocated memory for direct API use\\
  \smallcode{unregister\_direct\\\_memory} & Unregister client-allocated memory for direct API use\\
  \smallcode{put\_direct} & Zero-copy large put operation using client provided memory \\
  \smallcode{get\_direct} & Zero-copy large get operation using client provided memory \\
  \smallcode{get\_direct\_offset} & Read sub-region of pool memory directly \\
  \smallcode{put\_direct\_offset} & Write sub-region of pool memory directly \\  
  \hline
  \smallcode{async\_put\_direct} & Asynchronous version of put\_direct \\
  \smallcode{async\_get\_direct} & Asynchronous version of get\_direct. \\
  \smallcode{async\_get\_direct\\\_offset} & Asynchronous version of get\_direct\_offset \\
  \smallcode{async\_put\_direct\\\_offset} & Asynchronous version of put\_direct\_offset \\
  \smallcode{check\_async\\\_completion} & Check for asynchronous operation completion \\
  \hline
\end{tabularx}
\caption{Advanced Bulk-transfer API}
\label{tab:clientbulkapi}
\end{centering}
\end{table}

\subsubsection{Direct Offset Operations}

For operations on large areas of memory the ability to perform
sub-region read/write operations is useful.  For this, MCAS provides
the \code{xxx\_direct\_offset} APIs (see Table~\ref{tab:clientbulkapi})
These functions allow direct read and write operations for a region of the value
space (associated with a key) that is defined by base offset and size pair.

\subsection{Primary Index Component}

MCAS uses a primary index to manage the mappings from key-to-value and
an optional secondary index for scanning of the key-space.

The primary index is provided as a \textit{storage engine} component
that provides broader services such as memory management.  It is a
pluggable component that implements are predefined interface
(\code{IKVStore}).  There are currently three storage engines included
in MCAS: \hstore, \hstorecc and \mapstore.

\hstore uses persistent memory for the main hash table and volatile
memory for the memory allocator itself.  Alternatively, \hstorecc uses
a persistent memory based memory allocator.  This allocator is slower
than its volatile memory counterpart, but it does not require
rebuilding after reset.

For DRAM-only scenarios, the \mapstore backend is available.  This is
based on a C++ STL \code{ordered\_set}.

\subsubsection{Memory Management}

\begin{figure}
\centering
\includegraphics[width=1.0\linewidth]{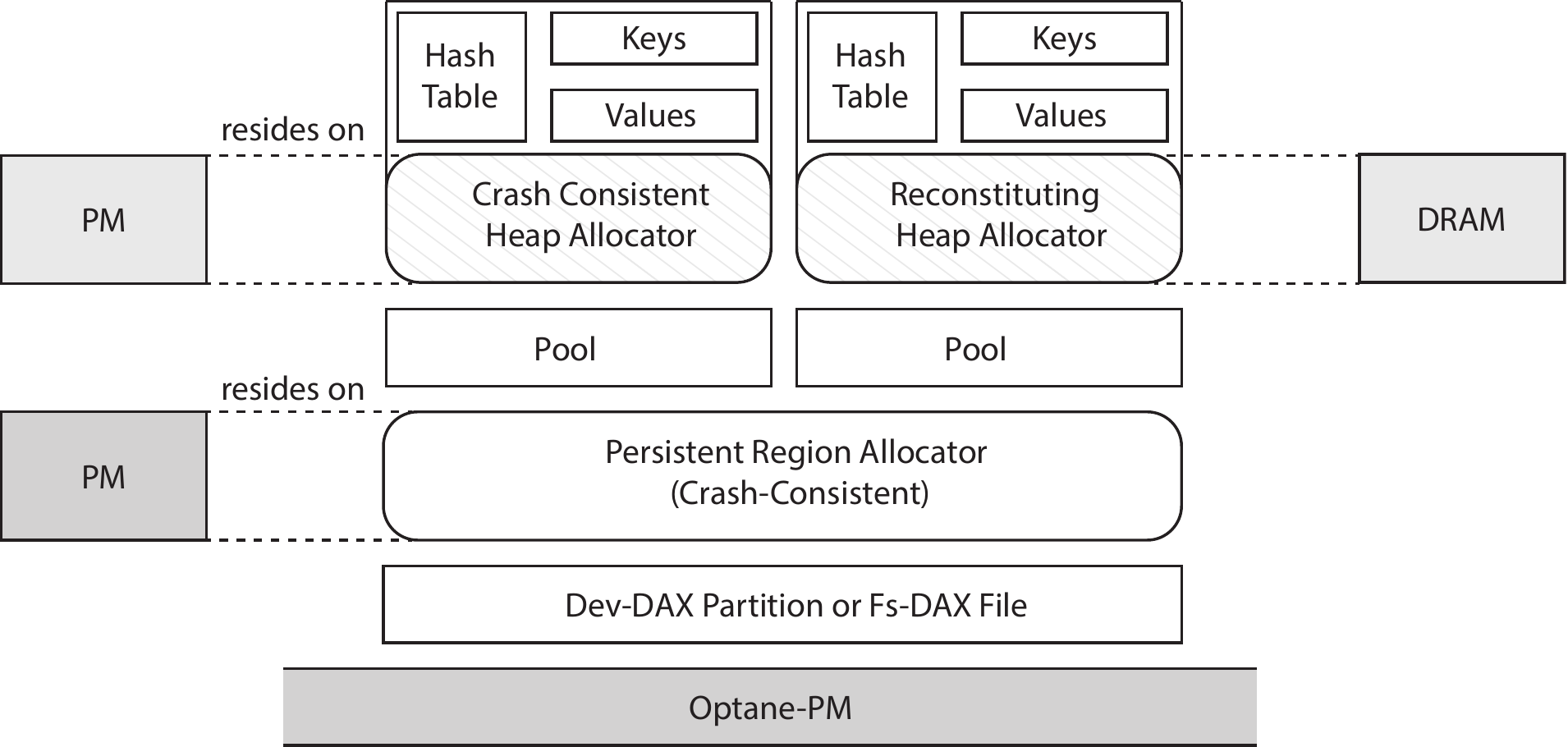}
\caption{Memory Management Architecture}
\label{fig:mem_arch}
\end{figure}

Figure~\ref{fig:mem_arch} provides an overview of the \hstore memory
management architecture.  At the lowest level, MCAS uses either a
\textit{device-DAX} partition, which is a fixed-size partition of a
given interleave set or an \textit{fs-DAX} file. Device-DAX is used
when On-Demand Paging (ODP) hardware is not available.  A device-DAX
partition or fs-DAX file is configured for each shard.

To manage the shard memory resources, \hstore and \hstorecc use a
coarse-grained crash-consistent heap allocator, known as the
\textit{region allocator}. This allocates memory for individual pools.
Because shards are inherently single-threaded the region allocator
need not be thread-safe and is therefore lock-less.  Memory is managed
at a 32MiB granularity using an array of region descriptors (offset,
length, identifier) that are maintained in persistent memory.  Updates
to region descriptors are made write-atomic by using a simple undo-log
approach.

MCAS has no restrictions\footnote{Actually, the size is currently
  restricted by the RDMA verbs maximum frame size, which is 1GiB.} on
key and value lengths.  This means that a heap-allocator is necessary
to support variable-length region allocation.  To support high-rates
of key-value pair insertion and deletion, \hstore maintains a heap
allocator for key-value data in volatile (DRAM) memory.  However,
because the state of the allocator is neither crash-consistent or
power-fail durable, we must ``reconstitute'' its state after restart.
To achieve this, MCAS uses the key-value length information that is
stored in \pmem (and is crash-consistent) to rebuild the allocation
state.

\begin{figure}
\centering
\includegraphics[width=1.0\linewidth]{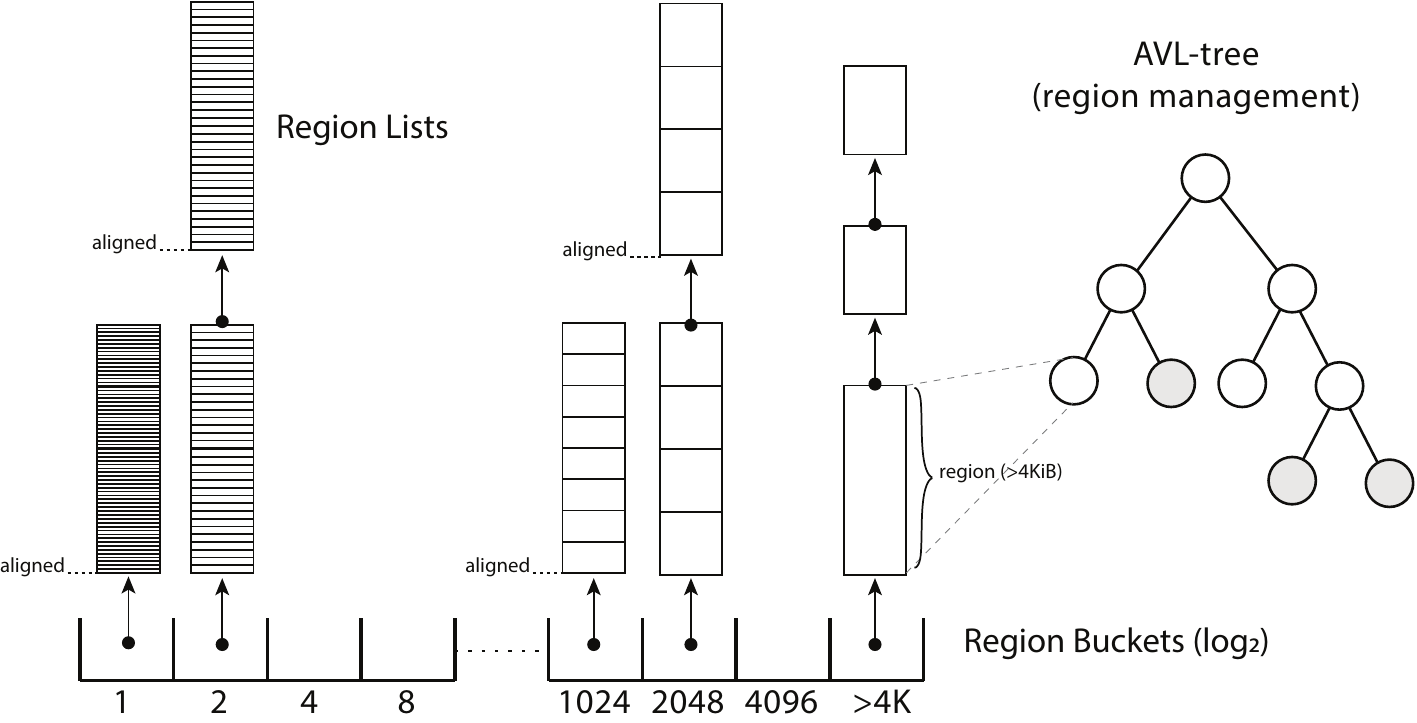}
\caption{Reconstituting Heap Allocator}
\label{fig:rcalloc}
\end{figure}

The reconstituting allocator manages a set of buckets that represent
different 2\textsuperscript{n} sizes (see Figure~\ref{fig:rcalloc}).
Objects (key or value data) up to a given size (4KiB) are allocated in
region lists.  Each region is 1MiB.  Objects are allocated by linear
scanning of the regions belonging to the corresponding bucket. If
there are no free slots in any region, a new region is allocated.
Regions and large-objects (> 4KiB) are allocated using an AVL-tree
based allocator.  If all slots in a region become free, the region can
be returned to the AVL-allocator.

\subsubsection{HStore Hash Table}

The core of \hstore is a hopscotch hash table~\cite{HerlihyST08}.  The
hash table is maintained in persistent memory. Expansion of the table
is achieved by adding successivevely larger \textit{segments}.  Each
additional segment doubles the table size.  Currently, shrinking of
the hash table is not supported.

\begin{figure}
\centering
\includegraphics[width=0.9\linewidth]{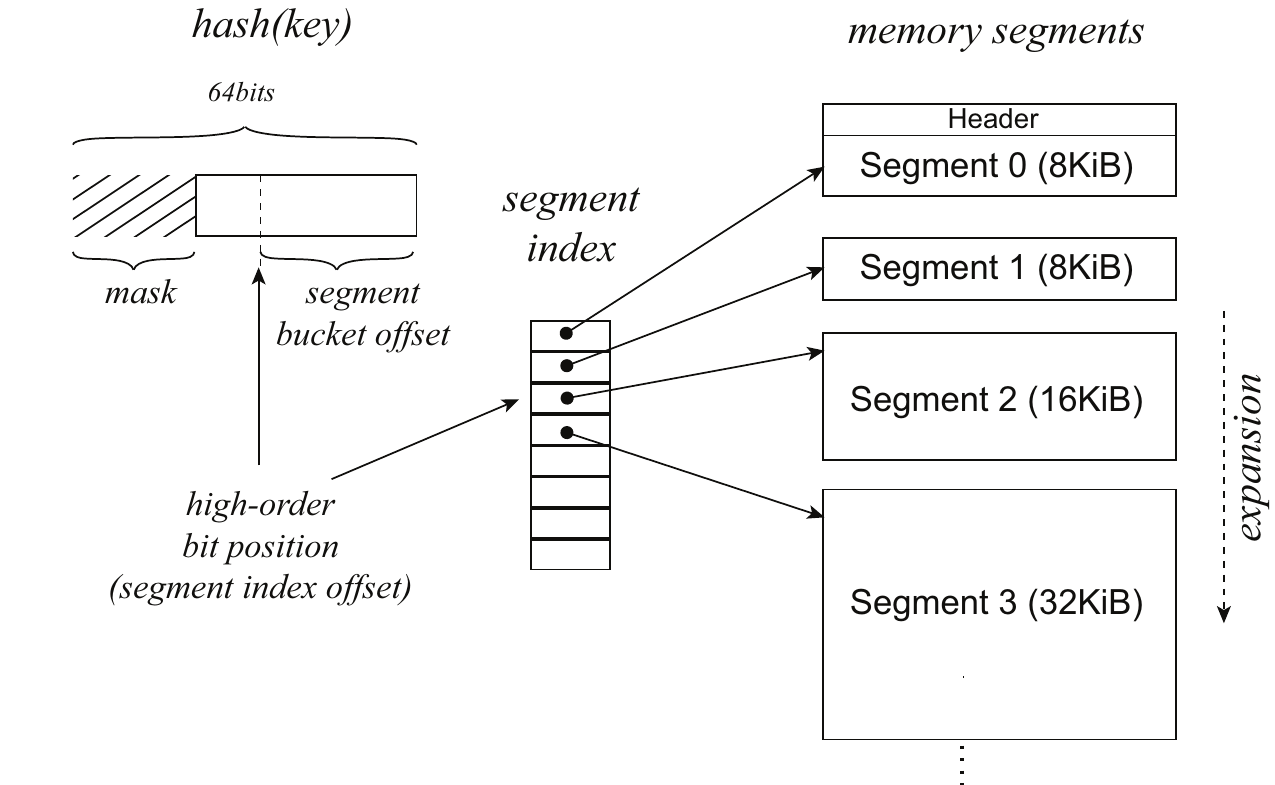}
\caption{Bucket-segment mapping}
\label{fig:buckseg}
\end{figure}

To manage the mapping between the 64-bit hashed key, we adopt a
strategy also used by the Intel TBB hash table
implementation~\cite{Pheatt:2008:ITB:1352079.1352134}; see
Figure~\ref{fig:buckseg}.  The basic idea is to partition the hash
value into a left hand mask (where bits are ignored) and a right hand
set of bits representing the segment index and the segment offset
(i.e.  designating the bucket).  The high-order bit, outside of the
mask, is used to indicate the segment index.  The remaining bits are
used to define the bucket/segment offset.

\begin{figure}[ht]
\centering
\includegraphics[width=1.0\linewidth]{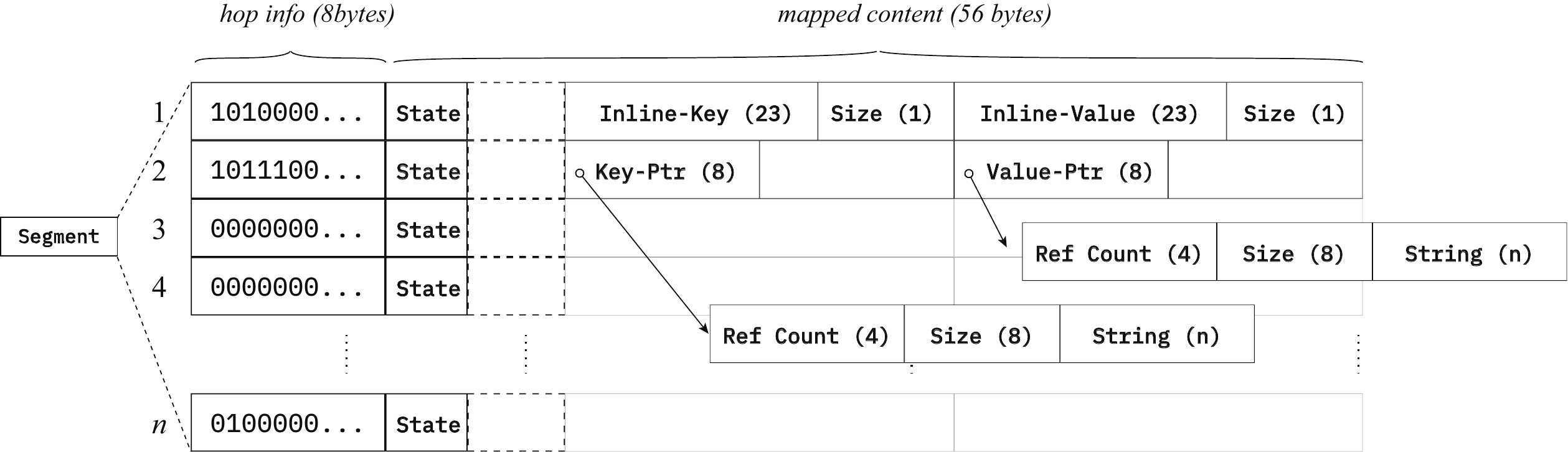}
\caption{Hash table arrangement}
\label{fig:hopscotch}
\end{figure}

The hash table memory layout is given in \ref{fig:hopscotch}.  Each
entry in the table contains a \textit{hop information} bitmap of \textit{H}
bits (\textit{H}=63), that indicates which of the next \textit{H} entries contain
items that hashed to the current entry's virtual bucket.  In addition
to the hop information, each entry contains a state field, key-value
pointer-size pairs, and in-lined keys and values (when sufficiently
small).  Each entry fits a single cache line (64 bytes).

Segments are added by reducing the left hand mask, thus enabling
another position for the high-order bit.  The segment index is needed
because new segments cannot be assured to be allocated in contiguous
memory.  They are linked together to aid navigation of buckets beyond
the segment boundaries.

\subsection{Secondary Index Component}

MCAS also supports the dynamic loading of a non-clustered ordered
index on the primary key, which we term the \textit{secondary
  index}\footnote{We use this term slightly different from the
  conventional database interpretation of forming an index on a
  different key from the primary key.}. This index, which is also
pluggable, implements a predefined interface (\code{IKVIndex}).  It
manages the key space only and its principal function is to provide an
index that can be efficiently scanned in key-order.  Scanning is based
on exact match, prefix or regular expression.  Currently, MCAS
provides only one (non-volatile) secondary index based on a volatile
red-black tree (C++ STL \code{map}).  Nevertheless, any alternative
secondary index can be easily developed and integrated into the
system.

\section{Active Data Objects (ADO)}
\label{sec:ado}

\begin{figure}
\centering
\includegraphics[width=0.8\linewidth]{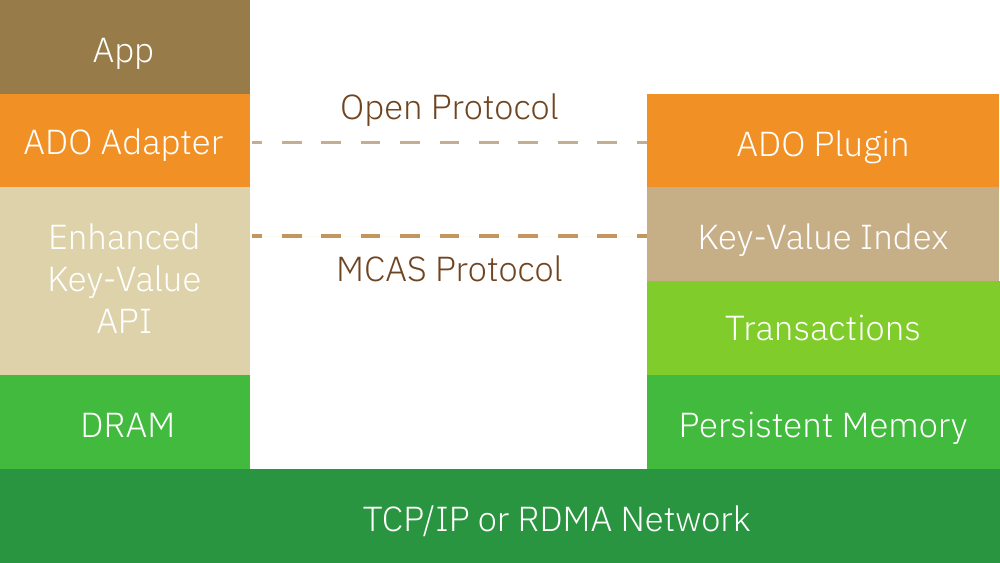}
\caption{ADO Open Protocol Architecture}
\label{fig:adoopenproto}
\end{figure}

A key differentiator for MCAS is its ability to perform ``push-down''
operations in what are termed Active Data Objects (ADO).  The ADO
mechanism is based on an \textit{open protocol} layering approach (see
Figure~\ref{fig:adoopenproto}) in which a developer can implement a
client-side library (adapter) and server-side plugin that handle and interpret a
custom protocol.  Together these two components are referred to as the
\textit{personality}.

The ADO plugin is statically associated with a shard through a
parameter in the MCAS configuration file.

\begin{minipage}{\linewidth}
\begin{lstlisting}[language=json,
    caption={Example MCAS two-shard configuration},
    captionpos=b, label={lst:adoconf}]
{
  "shards" :
  [
    {
      "core" : 0,
      "port" : 11911,
      "net"  : "mlx5_0",
      "default_backend" : "hstore",
      "dax_config" : [{
          "path": "/dev/dax0.0",
          "addr": "0x9000000000" }],
      "ado_plugins" : [
        "libcomponent-adoplugin-rustexample.so",
        "libcomponent-adoplugin-passthru.so"
      ],
      "ado_cores" : "2",
      "ado_params" :  {
        "param1" : "some param",
        "param2" : "and another"
      }
    }
  ],
  "ado_path" : "/mcas/build/dist/bin/ado",
  "net_providers" : "verbs"
}
\end{lstlisting}
\end{minipage}

Personalities can be used to create layered services and
functionality, both common and domain-specific.  Example common
services include replication, encryption, erasure-coding, versioning,
tiering, and snapshots.  Domain-specific services are centered around
data types, e.g., matrix operations, custom indices, and regular
expression matching.  Protocols are typically defined using Google
Flatbuffers, but other RPC frameworks can be used.

\begin{table}[ht]
\begin{centering}
\begin{tabularx}{1.0\linewidth}{
>{\setlength{\hsize}{.3\hsize}\raggedright\footnotesize}X
>{\setlength{\hsize}{.7\hsize}\raggedright\arraybackslash\footnotesize}X }
  \hline
  \textbf{Function} & \textbf{Description} \\
  \hline
  \smallcode{invoke\_ado} & Invoke Active Data Object \\
  \smallcode{async\_invoke\_ado} & Asynchronous version of invoke\_ado \\
  \smallcode{invoke\_put\_ado} & Invoke ADO with implicit put operation \\
  \smallcode{async\_invoke\_put\_ado} & Asynchronous version of invoke\_put\_ado \\
  \hline
\end{tabularx}
\caption{Advanced Client API}
\label{tab:clientadvapi}
\end{centering}
\end{table}

\subsection{ADO Invocation} \label{subsec:ADO Invocation}

To support the exchange of messages from the client to the ADO, MCAS
provides additional ``invocation'' APIs as summarized in
Table~\ref{tab:clientadvapi}.  These APIs are directed at a key-value
pair in an open pool.  In addition to the key, the parameters include
an opaque request, which encapsulates the protocol message:

\begin{figure}[h]
\begin{minipage}{\linewidth}
\begin{lstlisting}[frame=none]
status_t invoke_ado(
 const IMCAS::pool_t pool,
 const std::string& key,
 const void* request,
 const size_t request_len,
 const ado_flags_t flags,
 std::vector<ADO_response>& out_response,
 const size_t value_size = 0);
\end{lstlisting}
\end{minipage}
\end{figure}

\begin{figure}[ht]
\centering
\includegraphics[width=1.0\linewidth]{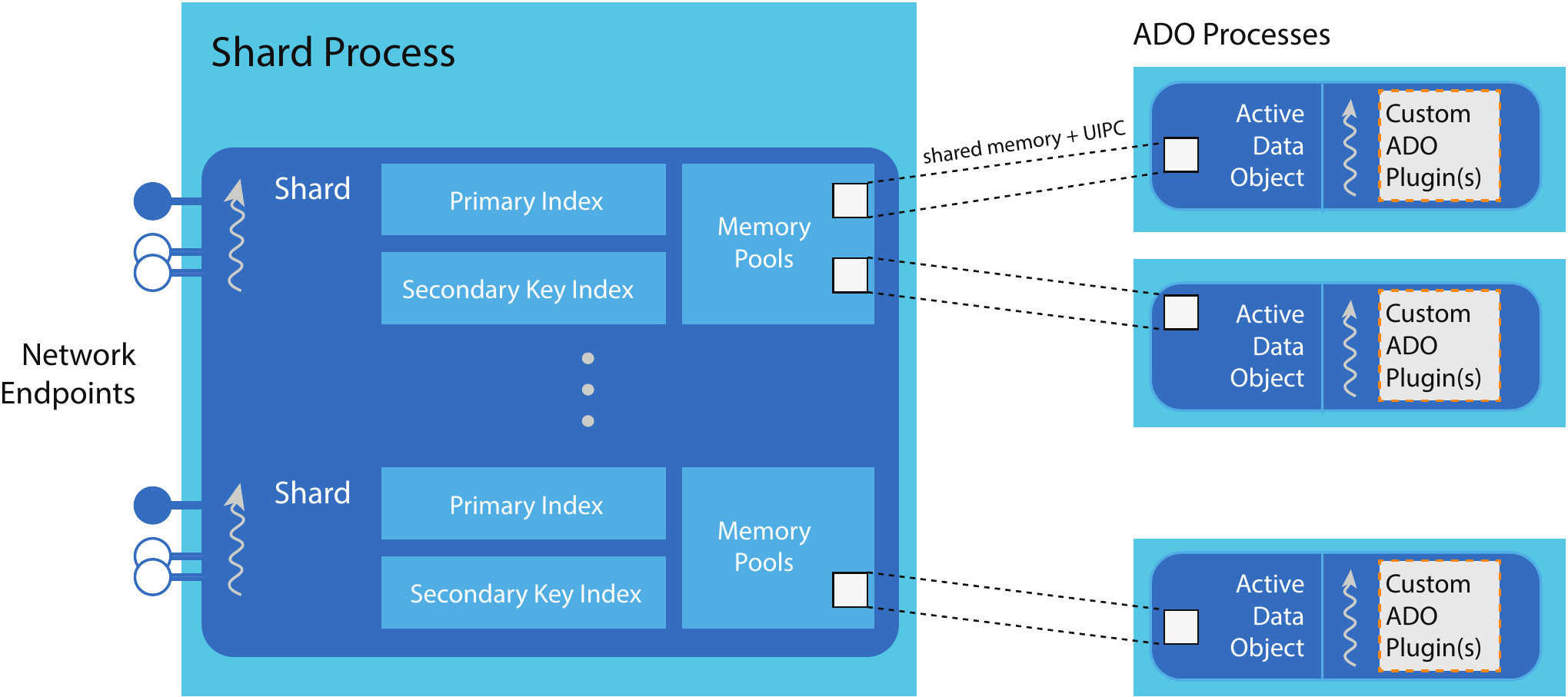}
\caption{ADO Architecture}
\label{fig:adoarch}
\end{figure}

The \code{invoke\_ado} invocation carries through to the ADO plugin on
Ethe server side.  Messages are passed from the main shard process to
the ADO process via a user-level IPC (UIPC) queue (see
Figure~\ref{fig:adoarch}).  UIPC is a user-level shared-memory region
that is instantiated with a lock-free FIFO.  Communications via UIPC
do not require a system call.  Before forwarding a message for an
\code{invoke\_ado} the shard thread locks the key-value pair so that
the ADO has the appropriate ownership.

On receiving a message from the shard (via the UIPC queue)
the ADO process calls the plugin-implemented \code{do\_work} method. This
method is defined as follows:

\begin{figure}[h]
\begin{minipage}{\linewidth}
\begin{lstlisting}[frame=none]
status_t do_work(
  const uint64_t              work_id,
  const char*                 key,
  const size_t                key_len,
  IADO_plugin::value_space_t& values,
  const void*                 in_work_request,
  const size_t                in_work_request_len,
  const bool                  new_root,
  response_buffer_vector_t&   response_buffers) = 0;
\end{lstlisting}
\end{minipage}
\end{figure}

Note that the \code{do\_work} upcall\footnote{Moving up the stack}
provides both key-value and request information as well as an indication of
whether the key-value pair has been newly created.  If multiple ADO plugins are
defined (see Listing~\ref{lst:adoconf}), they are called on a
round-robin schedule.  Responses from the work is collected as a
vector, which is ultimately passed back to the client.

On returning from the \code{do\_work} call, the ADO container process
returns a message to the shard thread via the UIPC.  On receipt of the
completion notification the shard thread releases the corresponding
lock.

MCAS also provides the \code{invoke\_put\_ado} variation in the API.
This variation allows a value to be \code{put} immediately prior to the
invocation/upcall of the ADO plugin (avoiding the need for the client
to perform two consecutive calls).

\subsubsection{Experimental Signaling Hooks}

There are occasions when an ADO requires notification of
non-ADO invoke reads and writes (e.g., \code{put}/\code{get}) to
a pool.  As an experimental feature, MCAS supports the configuration
of \textit{ado signals} that relay notifications from non-ADO shard
operations.

\begin{lstlisting}[language=json,
    caption={Shard-level ADO Signal Configuration},
    captionpos=b, label={lst:adosignalconf}]
  ...
  "ado_signals" : ["post-put", "post-erase"],
  ...
\end{lstlisting}

Signals are dispatched to the ADO via the UIPC queue. Before the UIPC
message is sent, the corresponding key-value pair is 'read' locked.
Signals propagate to the ADO plugin via a \code{do\_work} invocation
with a message prefix of `ADO::Signal'.  Responses to the clients are
postponed until completion of the ADO \code{do\_work} operation
(i.e. the client is stalled).

\subsection{ADO Execution Isolation}

The ADO plugin manipulates data that is
stored in persistent memory.  The shard process exposes the subject
pool memory to a separate ADO process.  This exposure is based either
on sharing of the fsdax file or performing a sub-space mapping with
devdax.  The latter requires a custom kernel module
(\code{mcasmod.ko}) to enable mapping between processes without a file
handle. Thus, the ADO process has complete visibility of the pool
memory, including the primary index; it cannot read or write memory
belonging to other pools.

ADO process instances effectively ``sandbox'' the compute performed by
the developer-supplied ADO plugin. The ADO process is launched on-demand and
remains operational while one or more clients has the corresponding
pool open.  Access to CPU cores and memory (persistent or DRAM) can be
specified via the MCAS shard configuration file. Operations performed
in the ADO generally cannot ``jam up'' or steal resources from the shard process.

Even though the ADO process has access to the primary index memory, it
does not manipulate this region of memory directly (avoiding thread
conflicts and the need for locking).  Instead, to perform operations
on the primary index, such as allocating a new key-value pair or
allocating memory from the pool, a callback interface is used.  The
callback functions are summarized in Table~\ref{tab:callbacks}.
Invocations on the callback interface by the plugin are passed back to
the shard thread via the UIPC queue.   

\begin{table}[ht]
\begin{centering}
\begin{tabularx}{1.0\linewidth}{
>{\setlength{\hsize}{.4\hsize}\raggedright\footnotesize}X
>{\setlength{\hsize}{.6\hsize}\raggedright\arraybackslash\footnotesize}X }
  \hline
  \textbf{Functions} & \textbf{Description} \\
  \hline
  \smallcode{create key/open key/erase key} & Key-value management \\
  \smallcode{resize value} & Resize existing value \\
  \smallcode{allocate/free memory} & Pool memory management \\
  \smallcode{get ref vector} & Get vector of key-value pairs \\
  \smallcode{iterate} & Iterate key-value pairs \\ 
  \smallcode{find key} & Scan for key through secondary index \\
  \smallcode{get pool info} & Retrieve memory utilization etc. \\
  \smallcode{unlock} & Explicitly unlock key-value pair \\
  \hline
\end{tabularx}
\caption{ADO plugin ``callback'' API}
\label{tab:callbacks}
\end{centering}
\end{table}

As previously discussed, locking for the ADO invoke target is explicitly
released when the call returns.  If the ADO operation creates or
opens other key-value pairs, then locks for these are also taken and
added to the deferred unlock list.

\subsubsection{Container-based Deployment}
MCAS is cloud-native ready.  Both the MCAS server process and ADO
processes can be deployed as Docker containers.  These containers are
built with provided Dockerfiles \code{Dockerfile.ado} and
\code{Dockerfile.mcas}.  Furthermore, MCAS can be deployed using the
Kubernates environment.  \code{mcas-server.yaml} is provided as a
reference pod configuration template.

\subsection{ADO Persistent Memory Operations}

Operations performed by the ADO that manipulate persistent memory must
be \textit{crash-consistent}.  That is, in the event of a
power-failure or reset, the data can be
``recovered'' to a known coherent state.  For example, given a list
insertion operation, in the event of recovery the list will either
have the element inserted or not at all; there will be no dangling
pointers or partially written data.

To support programming of crash-consistent ADO operations uses
memory-transactional programming through modified standard C++
templates libraries\footnote{We use EASTL from Electronic
Arts  because of its support to reset memory allocators.
(https://github.com/electronicarts/EASTL)}. 
The basic idea is to isolate the heap
memory for a data structure and instrument class methods with
\textit{undo logging}.  Each memory write is preceded by a copy-off
operation that saves the original state of the memory that will be
modified by the invocation. The copy-off operation is itself also
atomically transactional.  On completion of a program transaction,
which is made up of one or more method invocations on the data
structure, a \code{commit} call is made to clear the undo log.  In the
event of recovery from a ungraceful power-fail or reset event the undo
log is checked.  If the undo log is not empty, the copied-off regions
are restored to the heap and then the log is cleared.  This
effectively ``rewinds'' the data structure state to before the failed
transaction. More detail is given by the example listed in Appendix B.

Alternatively, PMDK or some other persistent programming methodology can
be used to write crash-consistent ADO operations. 


\section{Clustering}

To support coordination between MCAS processes at the ADO-level, MCAS
provides basic clustering support through the
\textit{libzyre}\footnote{https://github.com/zeromq/zyre} framework.
This library, based on ZeroMQ\footnote{https://zeromq.org/}, provides
proximity based peer-to-peer networking either through UDP beaconing
or a gossip protocol.  It uses reliable Dealer-Router pattern for
interconnections, assuring that messages are not lost unless a peer
application terminates.

Clustering is enabled by adding a \textit{cluster} section in the MCAS
configuration file, such as follows:

\begin{minipage}{\linewidth}
\begin{lstlisting}[language=json,caption={MCAS Process-level Cluster Configuration},captionpos=b, label={lst:clusterconf}]
{
    "cluster" :
    {
        "name" : "server-0",
        "group" : "MCAS-cluster-0",
        "addr" : "10.0.0.101",
        "port" : 11999
    },    
    "shards" :
    [
        {
          ...
\end{lstlisting}
\end{minipage}

Clustering is handled on a separate network port. Events, such
as a node joining or leaving a cluster, are propagated to each of
the shard threads and then to any active ADO processes. The plugin
method \code{cluster\_event} is up-called by the ADO thread receiving
notifications via the IPC queue:

\begin{minipage}{\linewidth}
\begin{lstlisting}[language=C++, frame=none]
void cluster_event(const std::string& sender,
                   const std::string& type,
                   const std::string& message);
\end{lstlisting}
\end{minipage}

Clustering can also be deployed directly in the ADO process directly
by using the cluster component (libcomponent-zyre.so).  This is a
useful approach when shard or pool activity needs to be associated to
a different cluster group name.





\section{Performance Evaluation}

In this section, we highlight the performance profile of MCAS through
experimental data. This results were collected from MCAS version v0.5.1
using the MPI-coordinated benchmark tool \code{mcas-mpi-bench}.

Our test system is based on the IBM Flashsystems platform,
which provides two PCIe-sharing canisters, each with two CPU sockets.
Table~\ref{tab:spec} gives the server system details.  To drive the
workload, four client systems are used. Even though the client systems are
of similar specification, they are not precisely the same.  This causes some
slight perturbation of results.

\begin{table}[ht]
\begin{centering}
\begin{tabularx}{1.0\linewidth}{
>{\setlength{\hsize}{.2\hsize}\raggedright\footnotesize}X
>{\setlength{\hsize}{.8\hsize}\raggedright\arraybackslash\footnotesize}X }
  \hline
  \textbf{Component} & \textbf{Description} \\
  \hline
  Processor & Intel Xeon Gold 5128 (Cascade Lake) 16-core 2.30GHz \\
  Cache & L1 (32KiB), L2 (1MiB), L3 (22MiB) \\ 
  DRAM & PC2400 DDR4 16GiB 12x DIMMs (192GiB) \\
  NVDIMM & Optane PMM 128GB 12x DIMMs (1.5TB) \\
  RDMA NIC & Mellanox ConnectX-5 (100GbE) \\
  OS & Linux Fedora 27 with Kernel 4.18.19 x86\_64 \\
  Compiler & GCC 7.3.1 \\
  NIC S/W & Mellanox OFED 4.5-1.0.1 \\   
  \hline
\end{tabularx}
\caption{Server system specification}
\label{tab:spec}
\end{centering}
\end{table}

\begin{figure}[]
\centering
\includegraphics[width=0.9\linewidth]{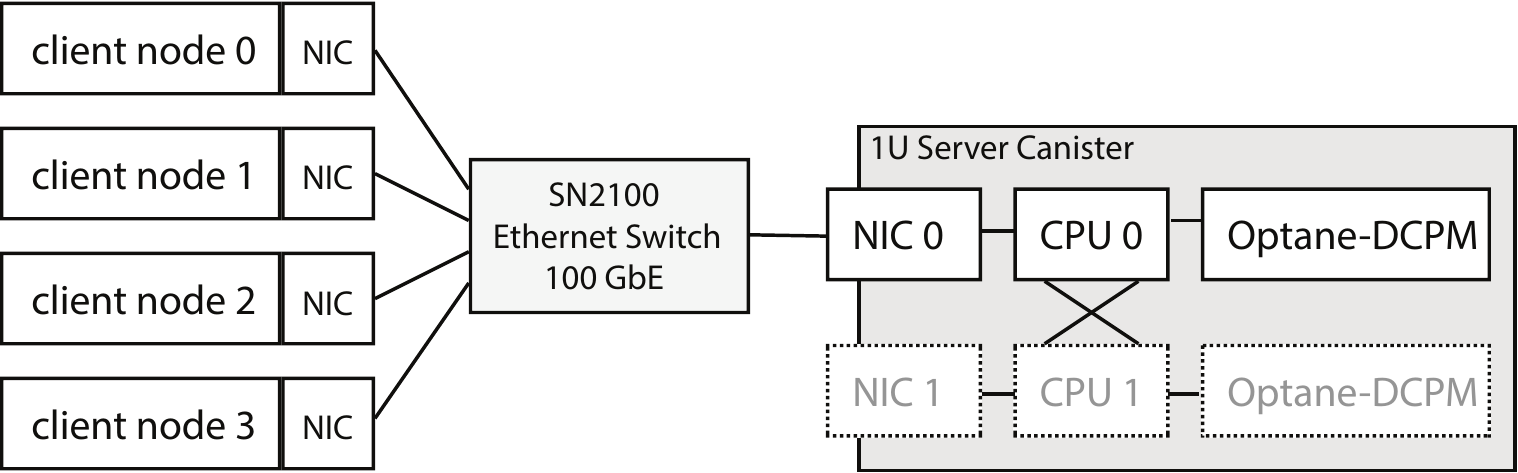}
\caption{Experimental Network Topology}
\label{fig:net_top}
\end{figure}

The server and client nodes
are connected to a 100GiB Ethernet Switch (see
Figure~\ref{fig:net_top}).

\subsection{Small Operations Throughput Scaling}

We examine the aggregate throughput of small \code{put} and \code{get}
operations (8-byte key, 16-byte value) with increasing number of
shards.  100\% read (\code{get}) and 100\% write (\code{put})
workloads are measured. Each shard is deployed as a separate process.
Five independent client threads, with separate pools and random keys,
drive the workload for each shard.  All client calls are synchronous
and the data is fully persistent and consistent on return.

\begin{figure}[ht]
\centering
\includegraphics[width=1.0\linewidth]{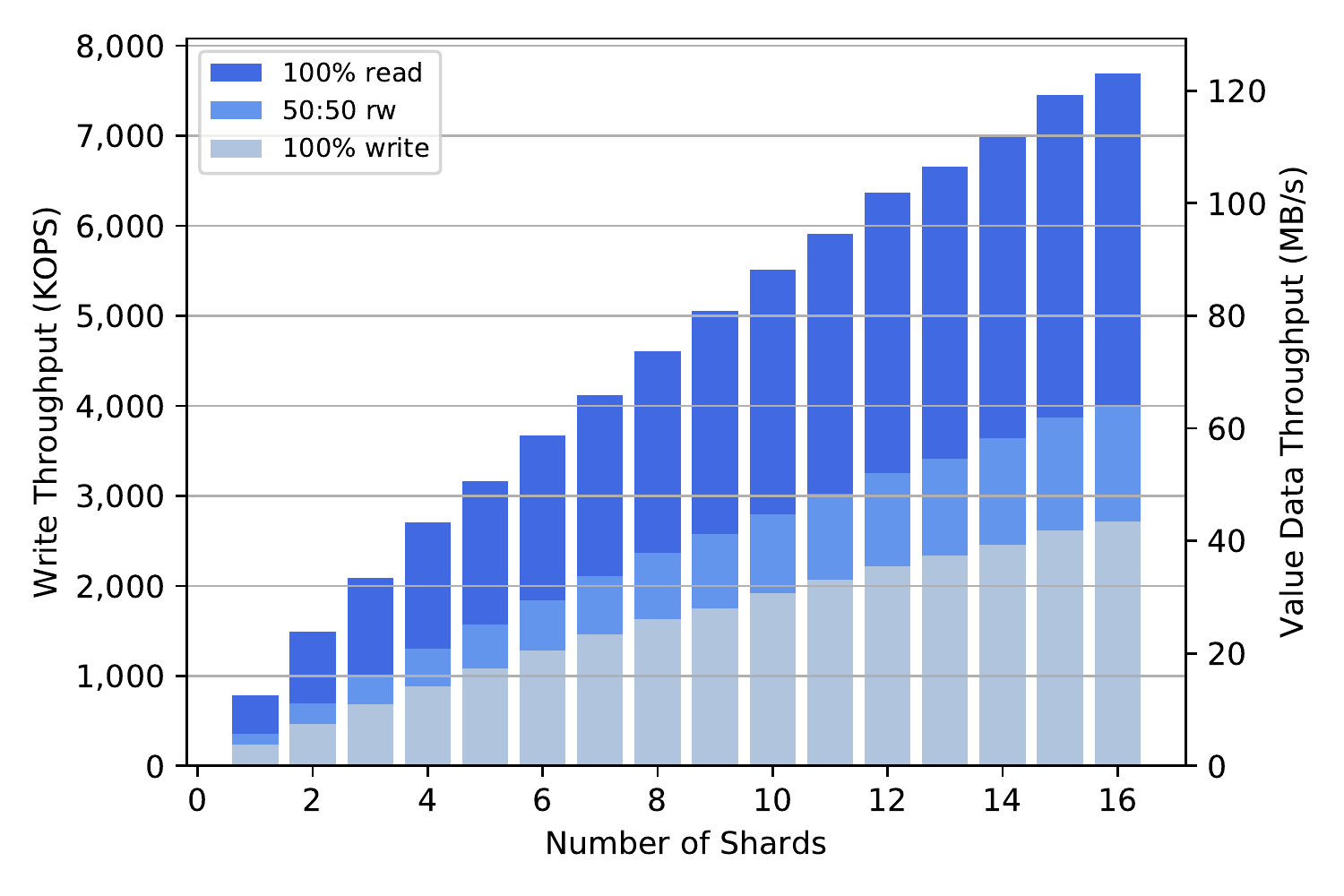}
\caption{Scaling of Small Operations}
\label{plot:smallop}
\end{figure}

The data shows that for 16 shards (i.e. a fully populated socket), the
system can support 2.72M puts/second and 7.69M gets/second.  At 16
shards, total degradation\footnote{Calculated from linear projection
  of observed single shard performance.} is ~39\% and ~29\%, for 100\%
read and 100\% write respectively.

\subsection{4KiB Operations Throughput Scaling}

We now examine throughput for \code{put} and \code{get} of larger 4KiB
values.  Key size remains at 8 bytes.  Note, this is not using the
zero-copy APIs (\code{put\_direct} and \code{get\_direct} that are
typically used for values above 128KiB.  The non-direct APIs must perform
memory copies on both client and server side.

\begin{figure}[ht!]
\centering
\includegraphics[width=1.0\linewidth]{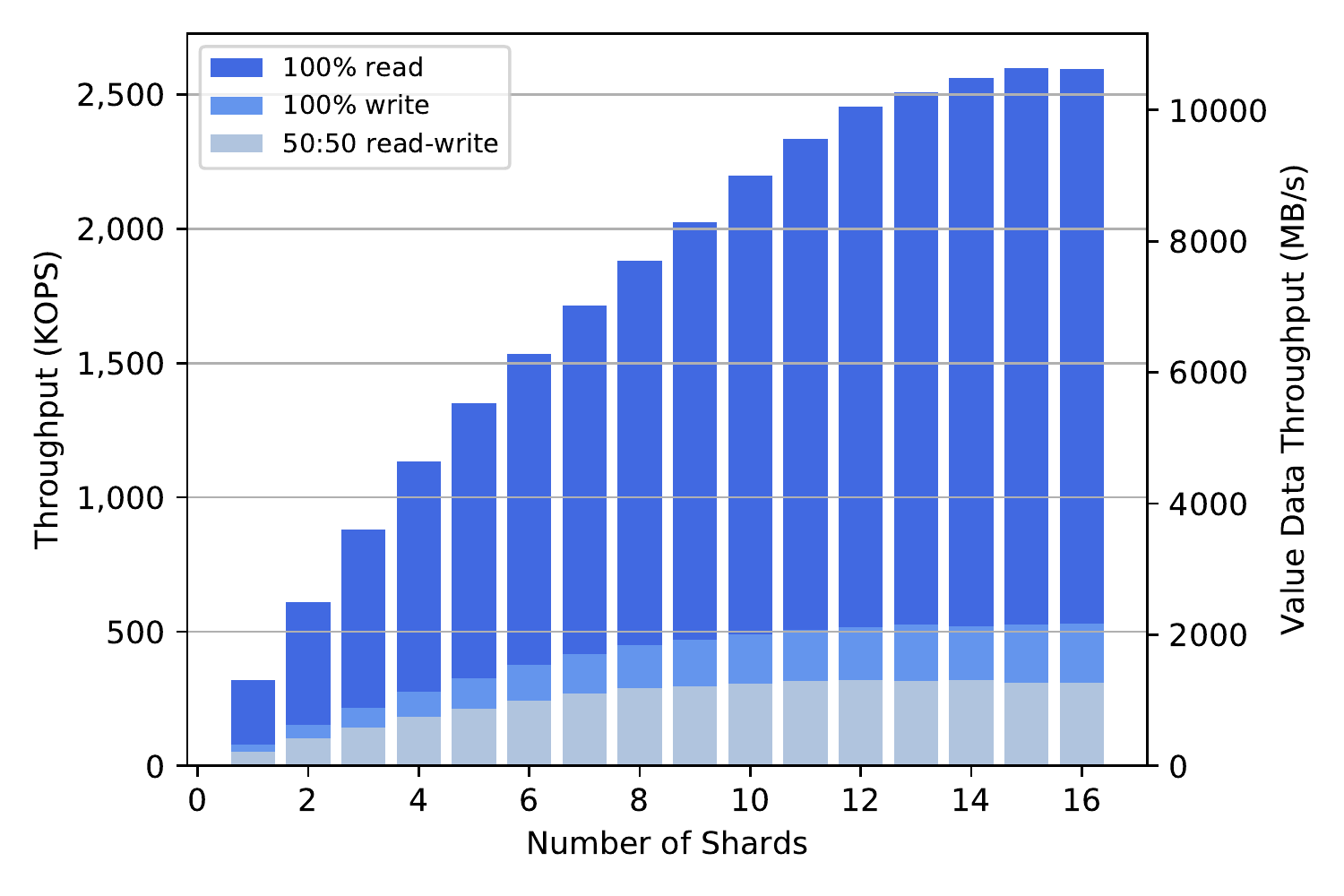}
\caption{Scaling of 4KiB Put/Get Operations}
\label{plot:4kop}
\end{figure}

Here read performance scales well up to around 12 shards.  Write performance
is considerably less. 

\subsection{Aggregate Scaling at Different Value Sizes}

\begin{figure}[ht!]
\centering
\includegraphics[width=1.0\linewidth]{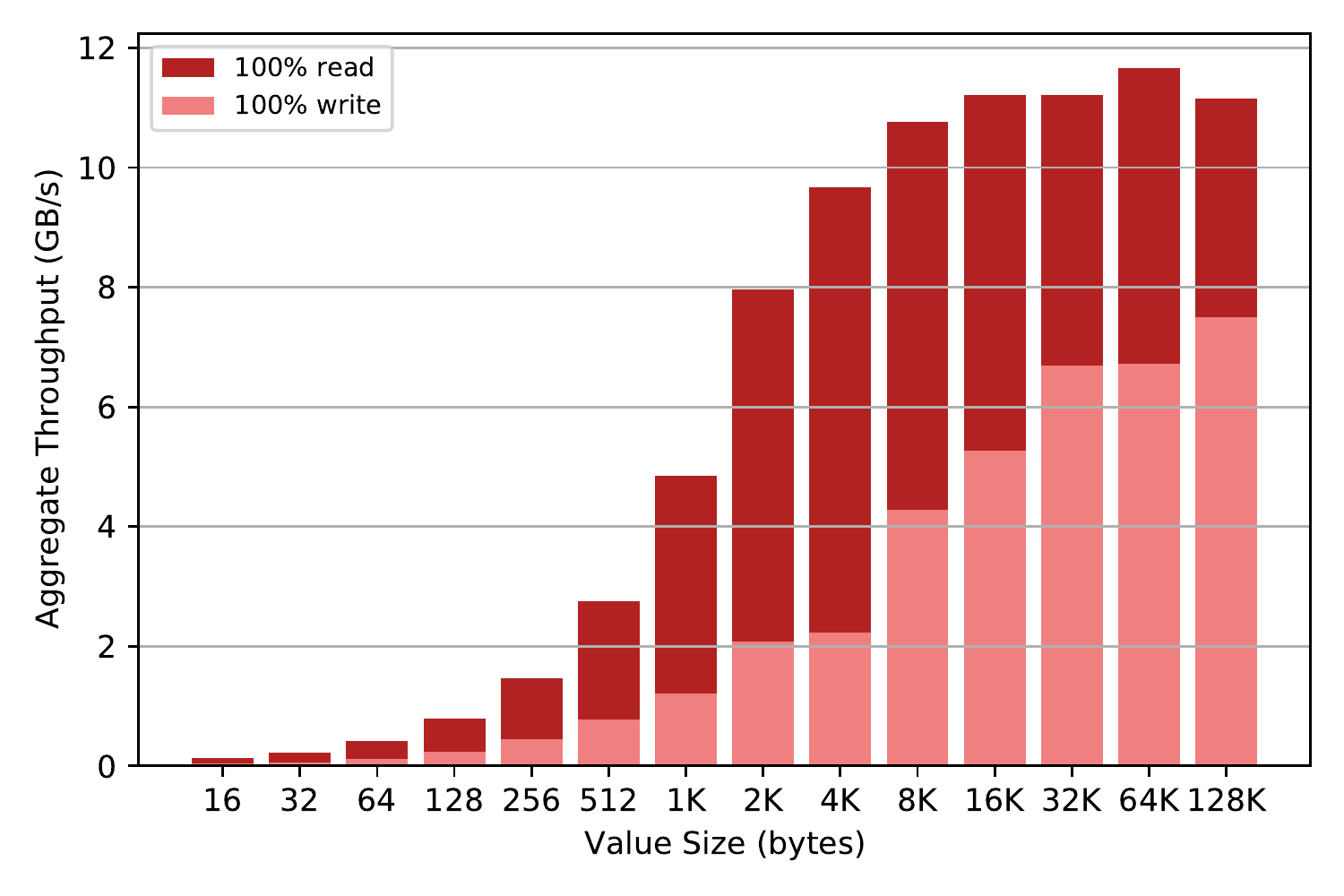}
\caption{Aggregate-Scaling of Throughput for Changing Value Size}
\label{plot:diffvalue-bw}
\end{figure}

We now examine aggregate throughput for 16 shards on a single socket
with network-attached clients (see Figure~\ref{fig:net_top}).
Figure~\ref{plot:diffvalue-bw} shows data for 100\% read and 100\%
write workloads.  Values are doubled in size from 16B to 128KiB. Note that
the operations are non-direct (i.e. copy based).

The results show that value sizes of 8KiB and above can saturate
over 90\% of the 100GbE bandwidth.

\subsection{128KiB Direct Operations Throughput Scaling}

We now examine performance scaling for the \code{put\_direct} and
\code{get\_direct} zero-copy operations.  Direct operations are not
used on smaller values due to overhead of performing scatter/gather
DMA.

For \code{get\_direct}, the 100GbE RDMA network connection becomes the
predominant limiter.  Network bandwidth is saturated at 6 shards.  For
\code{put\_direct}, the persistent memory is the limiter (at just over
10GiB/s).  This is congruent with findings reported
in~\cite{izraelevitz2019basic}.

\begin{figure}[ht!]
\centering
\includegraphics[width=1.0\linewidth]{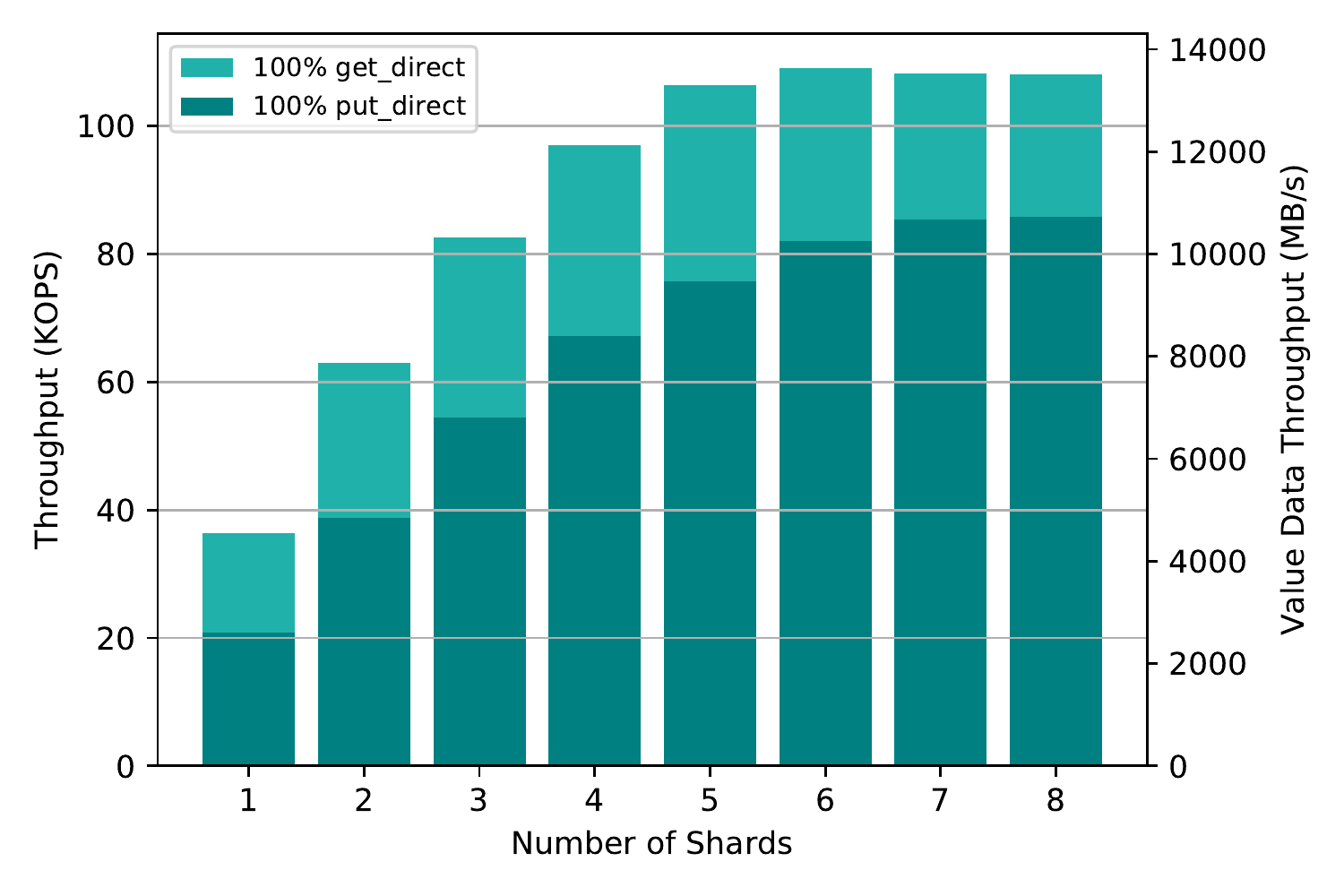}
\caption{Shard-Scaling of 128KiB Direct Put/Get Operations}
\label{plot:128kop}
\end{figure}

\goodbreak
\subsection{Latency}

A key differentiator for Persistent Memory is the ability to
achieve high throughput with low latencies.  The data is provided
as a histogram for 1 million samples across 40 bins.  Note that
the y-axes are presented with a logarithmic scale.

\begin{figure}[ht!]
  \centering

  \subfloat[Get (100\% read)]{
    \includegraphics[width=1.0\linewidth]{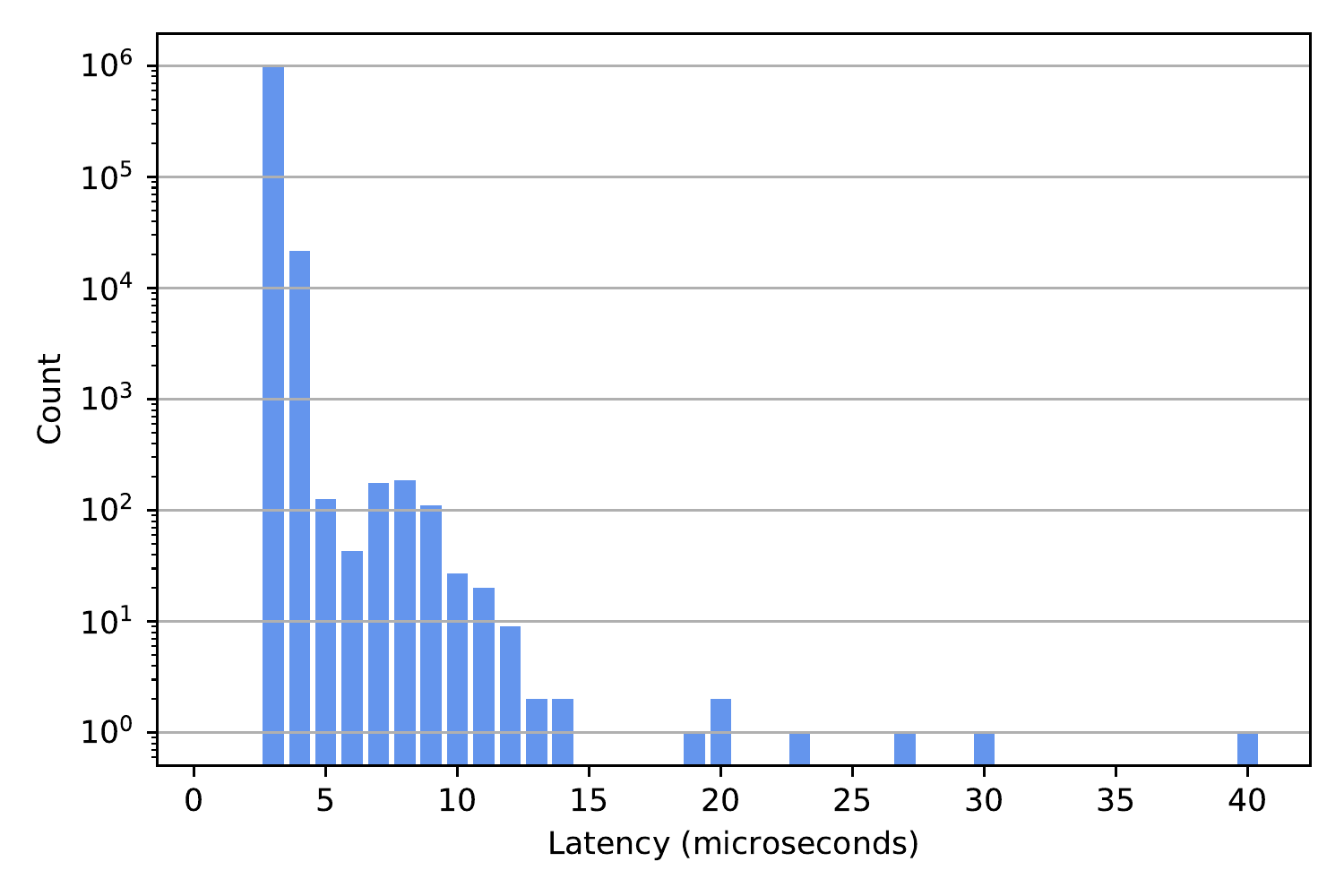}
  }

  \subfloat[Put (100\% write)]{
    \includegraphics[width=1.0\linewidth]{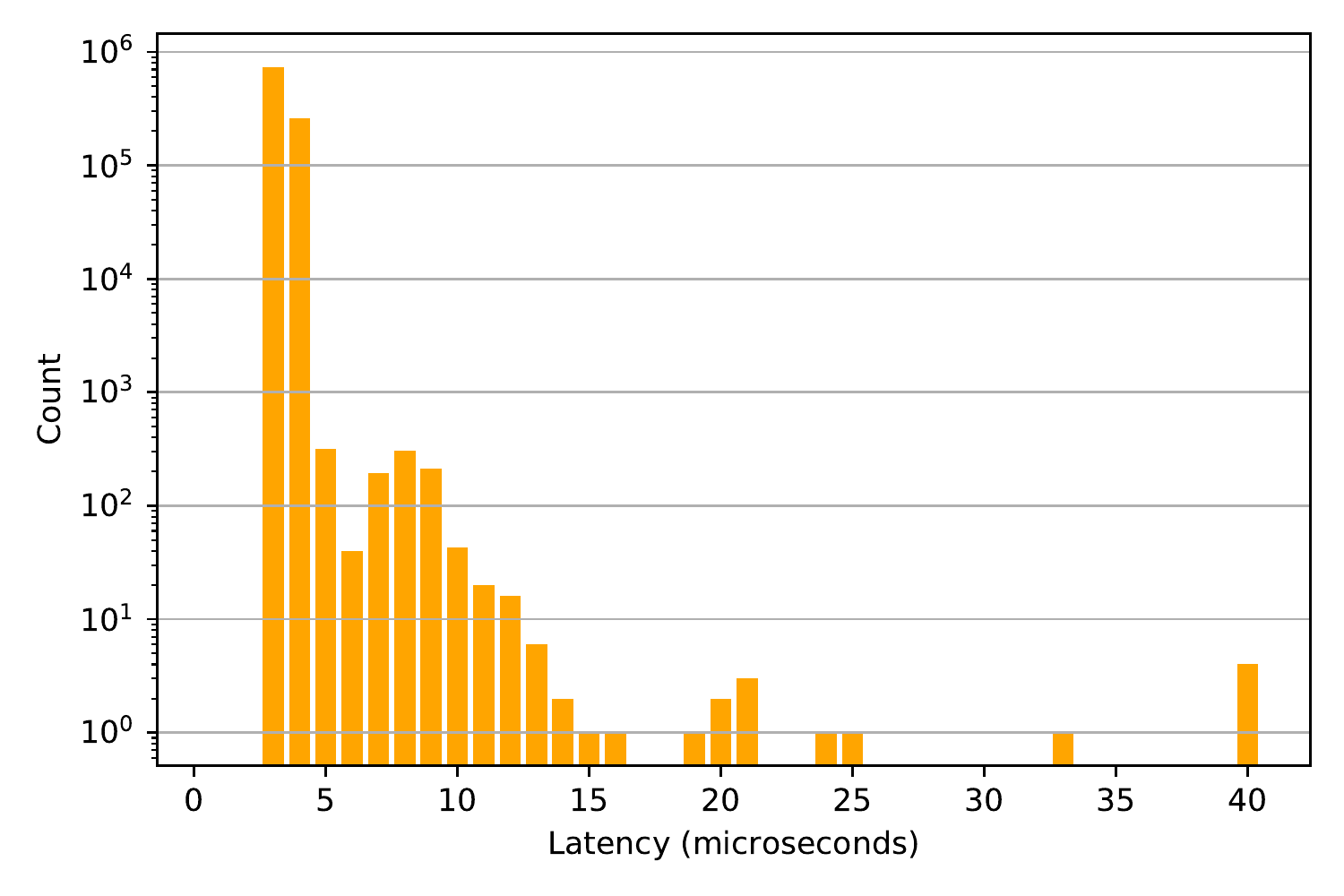}
  }

\caption{Latency Distribution for 16B Operations}
\end{figure}

\begin{figure}[ht!]
  \centering

  \subfloat[Get (100\% read)]{
    \includegraphics[width=1.0\linewidth]{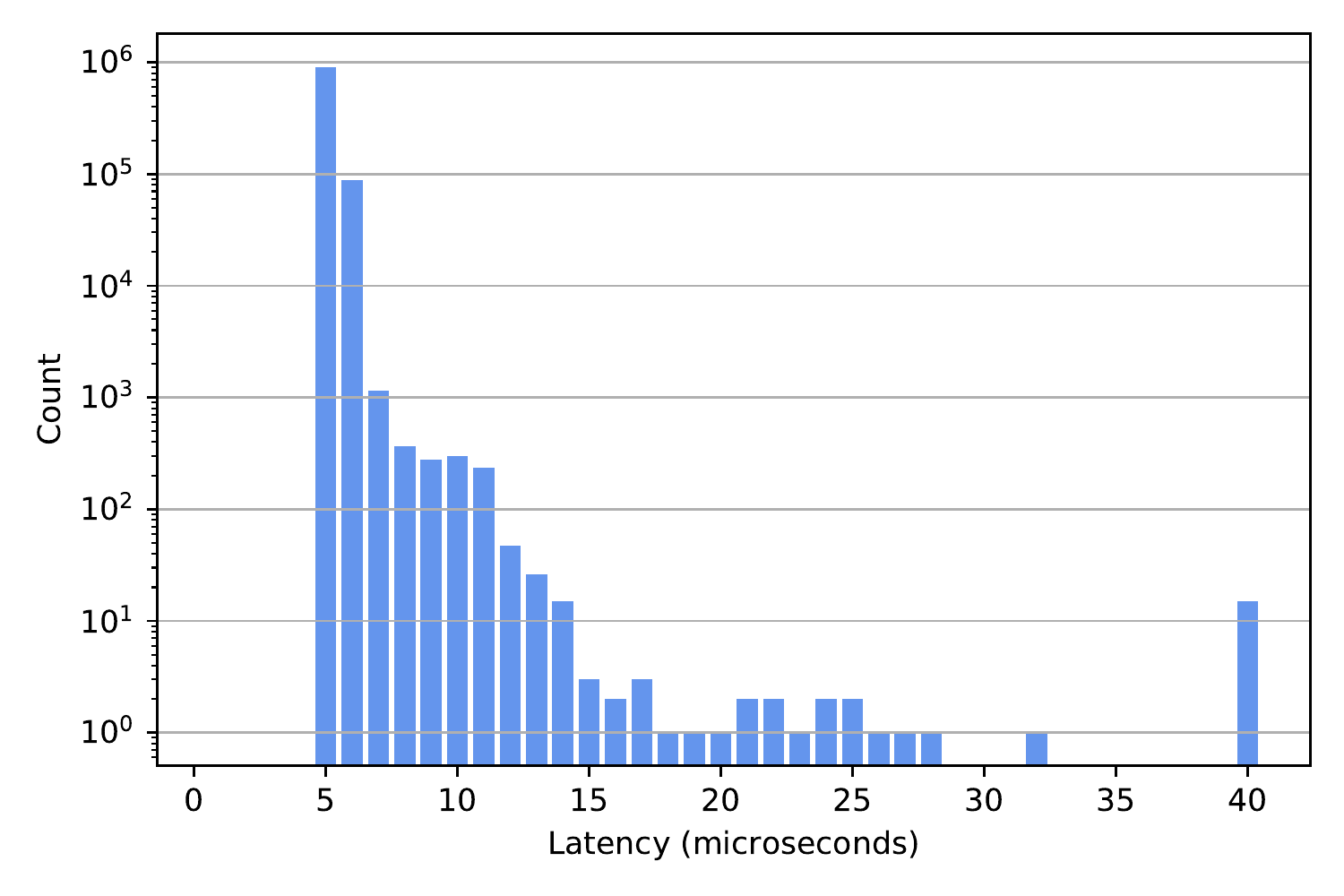}
  }

  \subfloat[Put (100\% write)]{
    \includegraphics[width=1.0\linewidth]{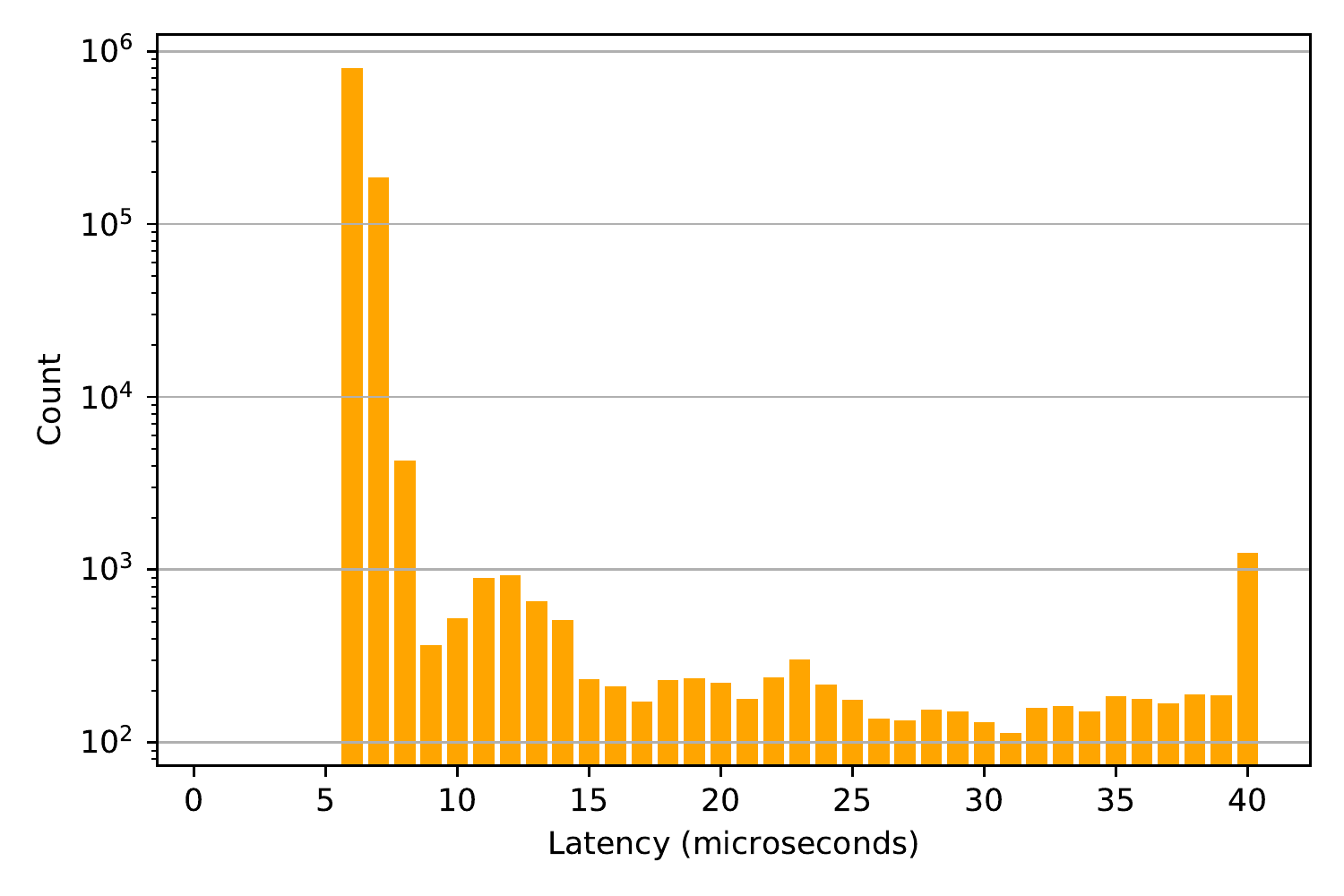}
  }
\caption{Latency Distribution for 4KiB Operations}
\end{figure}

\begin{figure}[ht!]

\centering
\includegraphics[width=1.0\linewidth]{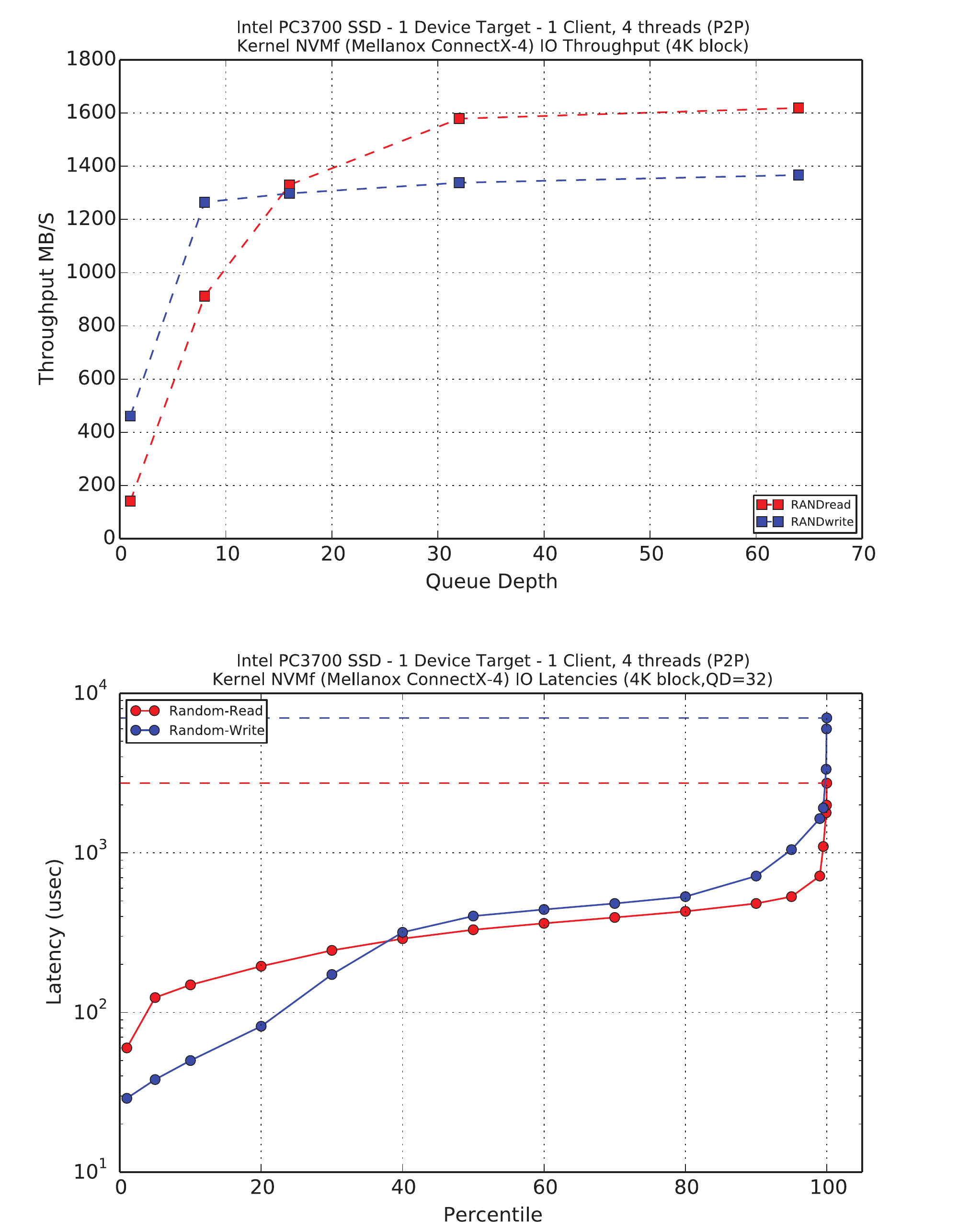}
\caption{Throughput-Latency Relationship for NVMe-over-fabric}
\label{fig:nvmef}
\end{figure}

To provide the reader some comparison with SSD NVMe,
figure~\ref{fig:nvmef} shows the relationship between throughput and
latency for NVMe-over-fabric (same network).

\newpage 
\subsection{Client Aggregation Fairness}

Here we look at how client performance is affected by ``competition''
on the same shard.  To do this, we collected throughput measures
(16-byte 100\% get operations) for an increasing number of threads.

\begin{figure}[ht!]
\centering
\includegraphics[width=1.0\linewidth]{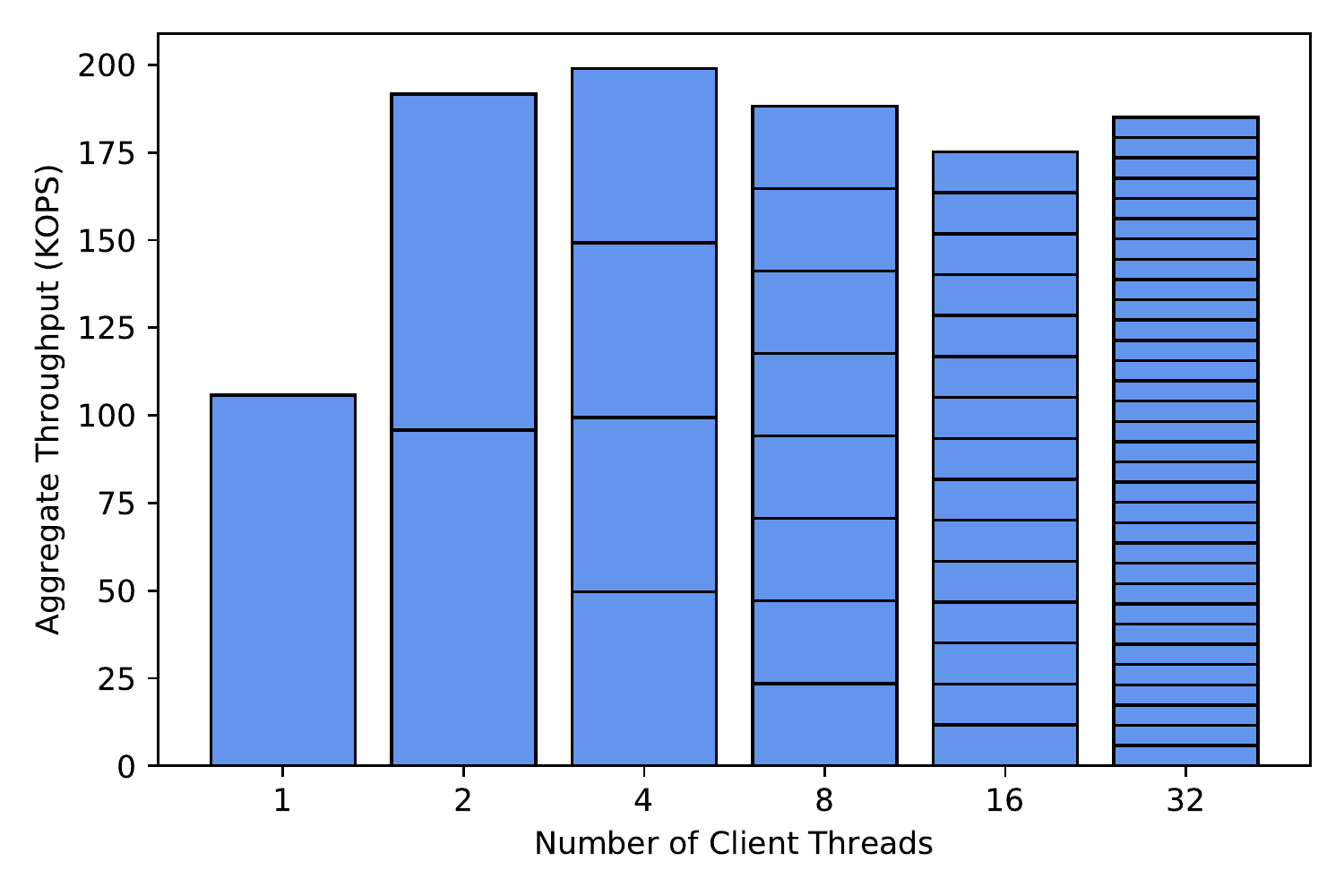}
\caption{Aggregation Behavior}
\label{plot:agg0}
\end{figure}

Saturation occurs at as little as four client threads.  Thereafter,
some nominal degradation occurs, but sharing is relatively fair
(i.e. each client gets its $n$-th share of performance where $n$ is
the number of clients).

\subsection{ADO Invocation Throughput Scaling}

To examine the throughput of ADO invocations in MCAS, we use the
\code{ado\_perf} tool. The test makes invocations using an 8-byte
message payload.  On the server, the shards are configured with the
``passthru'' plugin (\code{libcomponent-adoplugin-passthru.so}) which
performs no real compute.  The results shown are for a single-key
target, and a pool of keys target.  In the latter, 100K keys are used
that belong to the same pool.

\begin{figure}[ht!]
\centering
\includegraphics[width=1.0\linewidth]{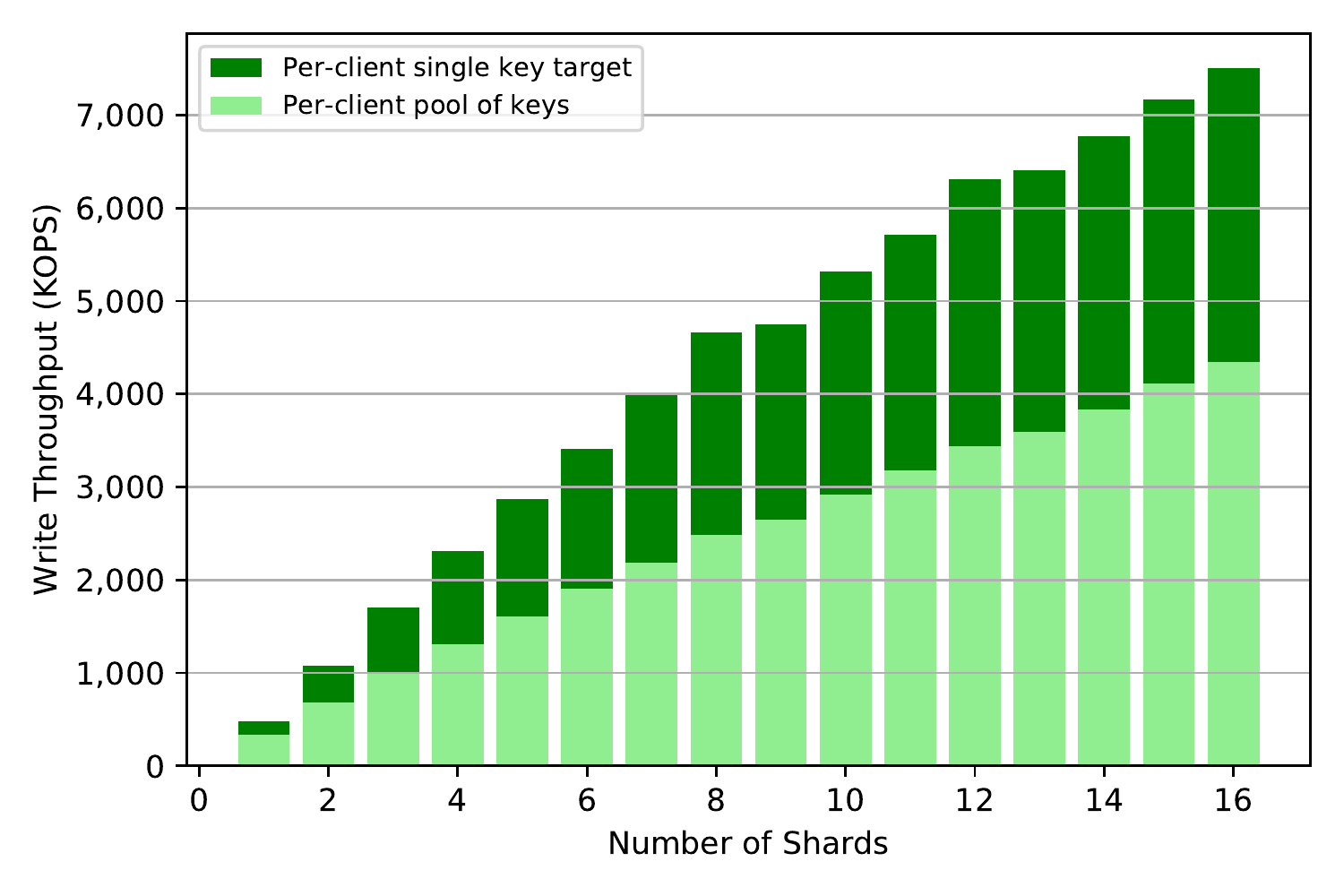}
\caption{Scaling of ADO Invocations}
\label{plot:adoscaling}
\end{figure}

The results show that at 16 shards, aggregate throughput is 7.5M IOPS
and 4.3M IOPS for same-key and key-set tests respectively.
Degradation from linear is 1.4\% for same-key and 18\% for key-set.

\section{Further Work}

Currently, MCAS is a research prototype. 
As we mature the platform to
a more robust production-grade solution, the following new features
will be considered:

\begin{itemize}
\item Additional language bindings (e.g., Rust, Go, Java).
\item Pool level authentication and admission control.
\item Improvements to crash-consistent ADO development (e.g., via h/w or compiler support).
\item Unified ADO protocol for enhanced key-value operations.
\item Boilerplate code for client-side replication.
\item Boilerplate code for FPGA accelerator integration.
\end{itemize}

\section{Availability}

The code for MCAS is available under an Apache 2.0 open source
license at https://github.com/IBM/mcas/.  Additional information is
also available at https://ibm.github.io/mcas/.

\section{Conclusion}

MCAS is a new type of system that aims to offer memory-storage convergence.  Based on
persistent memory and RDMA hardware, MCAS extends the traditional key-value paradigm
to support arbitrary (structured) value types with the potential for in-place operations.
Our initial experiments show that the ADO approach can result in a significant reduction
in network data movement and thus overall performance.

Going forward, our vision is to take MCAS beyond persistent memory and
position it for emerging Near-Memory Compute (NMC) and
Processing-in-Memory (PIM)
technologies~\cite{10.1145/2845084,10.1145/3299874.3317977,10.1145/3307650.3322237,
  10.1145/3036669.3038242, 10.1145/3400302.3415772, 10.1145/2997649}.
These technologies will be vital in addressing the memory-wall and
processor scale-up problems.
\bibliographystyle{abbrv} 
\bibliography{references}


\appendix
\onecolumn 
\section*{Appendix}

\subsubsection*{A. MCAS protocol definition}

\begin{figure}[h]
\centering
\includegraphics[width=0.9\linewidth]{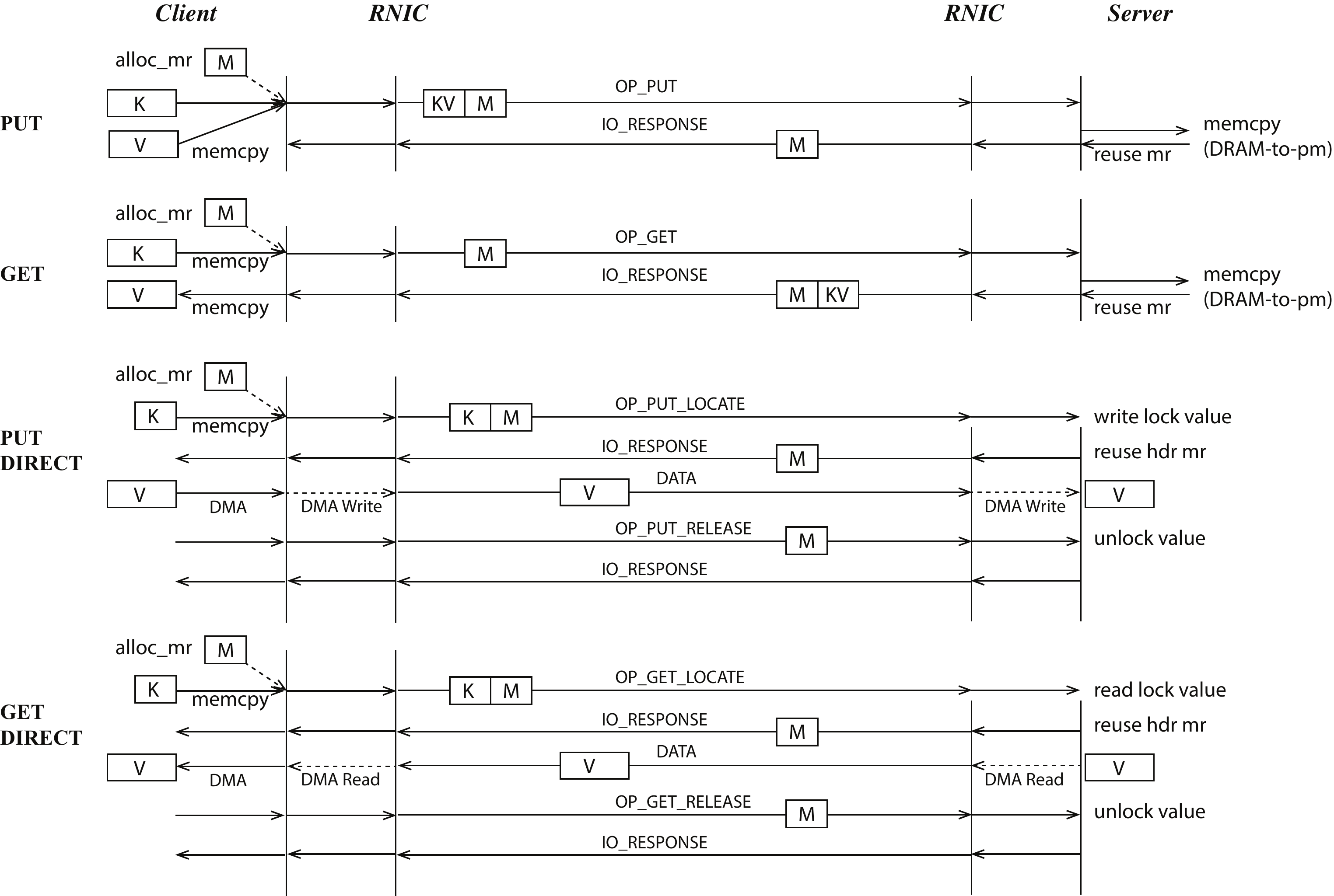}
\caption{MCAS Protocol}
\label{appendix:protocol}
\end{figure}

\newpage
\subsubsection*{B. Example C++ template-based crash-consistent programming}
\label{appendix:ccpm}

\begin{lstlisting}[frame=none]
/* define crash-consistent bitset type */      
using cc_bitset = ccpm::container_cc<eastl::bitset<52,
                                                   std::uint64_t,
                                                   ccpm::allocator_tl::tracker_type>>;

/* (re-)establish memory for crash-consistent heap allocator */                                                   
auto heap_regions = ccpm::region_vector_t(heap_area, heap_size);

ccpm::cca * cca; /* pointer to crash-consistent allocator */
cc_bitset * ccbitset; /* pointer to bitset data structure */


if (new_root) { /* new_root == true when key-value is first instantiated */
  /* create new allocator */
  cca = new ccpm::cca(heap_regions);
  /* create new bitset instance and pass in allocator */
  ccbitset = new (cca->allocate_root(sizeof(cc_bitset))) cc_bitset(*cca);
  ccbitset->commit();
}
else {
  /* reconstitute allocator */
  cca = new ccpm::cca(heap_regions, ccpm::accept_all);
  /* cast re-based root pointer */
  ccbitset = reinterpret_cast<cc_bitset*>(::base(cca->get_root()));
  ccbitset->rollback();
}

/* make changes to bitset data structure */
ccbitset->container->flip(4);
ccbitset->container->flip(5);
ccbitset->commit();

/* or make changes and roll back */
ccbitset->container->flip(6);
ccbitset->rollback();
\end{lstlisting}

\newpage
\subsubsection*{C. Example using ADO for multi-versioning}
\label{appendix:ado_versioning}
In this section, we describe a simple ADO example that adds versioning
to the basic key-value store.
\footnote{https://github.com/IBM/mcas/blob/master/examples/personalities/cpp\_versioning/}
The ADO creates an area in persistent memory that saves multiple
versions of values for a specific key and allows the client to
retrieve prior versions of a value.  In the ADO layer we ``raise''
the \code{get} and the \code{put} operations into the ADO handling
above the basic client API (see Table~\ref{tab:clientapi}). The code
is split into client-side library and server-side plugin. The message
protocol implementation is based on flatbuffers.

A client can invoke \code{put} and \code{get} operations (see Code
Listing~\ref{lst:client versioning}).  Under the hood, \code{put}
and \code{get} invocations result in calls to \code{invoke\_put\_ado}
and \code{invoke\_ado} respectively.

Corresponding messages are
constructed and sent as part of the ADO invoke payloads.  In both operations, the target is a
specific pool and key pair.  The messages are transmitted from the
client over the network to the main shard process and then transferred to
the ADO process via user-level IPC (see
Section~\ref{subsec:ADO Invocation} for more detail).
  
The ADO plugin handling starts with an up-call to the \code{do work}
 function (see Code Listing~\ref{lst:server_versioning}).  Here, the
 message is unpacked and then dispatched to the appropriate \code{put}
 or \code{get} handler.  The root pointer for the data structure that
 handles the different versions is provided as part of the \code{do
 work} invocation.  If it is the first-ever creation of the key the
 ADO plugin must initialize the root data structure.  In this example,
 the versioning metadata operations are made crash-consistent by using
 a basic undo log to ensure power-fail atomicity. In this example, we
 are using the \code{pmemlib} library and explicit 64-bit
 transactions\footnote{Building handcrafted crash-consistency can be
 complex but there are frameworks such as PMDK that can help to write
 crash-consistent code}.

During \code{put} invocation, the ADO handler creates (and persists)
an undo log that records the prior value and then perform a
transaction.  On successful completion of the transaction, the undo
log is cleared.  When the system starts the ADO checks for the need to
recover.  If the undo log is not clear, the logged data is copied back
to the original location in memory.  Finally, the result of \code{put}
and \code{get} operations are packed into a flatbuffer message and
returned to the client.

\newpage

\begin{lstlisting}[caption={Client-side for ADO versioning},
captionpos=b, label={lst:client versioning}, 
frame=none]
status_t Client::put(const pool_t pool,
                     const std::string& key,
                     const std::string& value)
{

  /* create request  message */
  ...
  s = _mcas->invoke_put_ado(pool,
                            key,
                            fbb.GetBufferPointer(),
                            fbb.GetSize(),
                            value.data(),
                            value.length() + 1, 
                            128, //root value length
                            component::IMCAS::ADO_FLAG_DETACHED,
                            response);
  return s;
}

status_t Client::get(const pool_t pool,
                     const std::string& key,
                     const int version_index,
                     std::string& out_value)
{

  /* create request  message */
  ...
  s = _mcas->invoke_ado(pool,
                        key,
                        fbb.GetBufferPointer(),
                        fbb.GetSize(),
                        0,
                        response);
  return s;
}

\end{lstlisting}

\begin{minipage}{\linewidth}
\begin{lstlisting}[caption={Server plugin for ADO versioning},
captionpos=b, label={lst:server_versioning}, 
frame=none]
status_t ADO_example_versioning_plugin::do_work(const uint64_t work_key,
                                        const char * key,
                                        size_t key_len,
                                        IADO_plugin::value_space_t& values,
                                        const void *in_work_request,
                                        const size_t in_work_request_len,
                                        bool new_root,
                                        response_buffer_vector_t& response_buffers)
{
  auto value = values[0].ptr;
  auto root = static_cast<ADO_example_versioning_plugin_root *>(value);
  if(new_root) {
    root->init();
  }
  else {
    root->check_recovery();
  }

  if(msg->element_as_PutRequest()) {
    ...
    //  Put
    auto value_to_free = root->add_version(detached_value, detached_value_len, value_to_free_len);
  /* create response message */
    return S_OK;
  }
  else if(msg->element_as_GetRequest()) {
    ...
    // Get
    root->get_version(pr->version_index(), return_value, return_value_len, timestamp);
  }
  /* create response message */
  ...
  return S_OK;
}

void init()
{
    pmem_memset_persist(this, 0, sizeof(ADO_example_versioning_plugin_root));
}

void check_recovery() {
    /* check for undo */
    if(_undo.mid_tx()) {
       /* recover from the undo log and then clear the undo log */
      _values[_current_slot] = _undo.value;
      _undo.clear();
    }
 }


void * add_version(void * value, size_t value_len, size_t& rv_len)
{
   void * rv = _values[_current_slot];
   rv_len = _value_lengths[_current_slot];

   /* create undo log for transaction */
   _undo = { _current_slot, _values[_current_slot], _value_lengths[_current_slot], _timestamps[_current_slot] };
   pmem_persist(&_undo, sizeof(_undo));

   /* perform transaction */
   _values[_current_slot] = value;

   pmem_persist(&_current_slot, sizeof(_current_slot));

   /* reset undo log */
   _undo.clear();
              
   return rv; /* return value to be deleted */
}

void get_version(int version_index, void*& out_value, size_t& out_value_len, cpu_time_t& out_time_stamp) const
{
   int slot = _current_slot - 1;
   while(version_index < 0) {
     slot--;
     if(slot == -1) slot = MAX_VERSIONS - 1;
     version_index++;
   }
   out_value = _values[slot];
}


\end{lstlisting}
\end{minipage}

\end{document}